\documentclass[11pt,a4paper]{article}

\usepackage{hyperref}

\usepackage[fleqn]{amsmath}
\usepackage{amsthm,amssymb,epsfig,latexsym,euscript}

\usepackage{pgf}
\usepackage{tikz}

\newtheorem{prop}{Proposition}[section]

{\theoremstyle{remark}
\newtheorem{rem}{Remark}[section]}
{\theoremstyle{remark}
}
\def\Proof{\medskip\noindent {\it Proof --- \ }}

\let\qed=\cqfd

\makeatletter
\@addtoreset{equation}{section}
\makeatother

\newcommand{\bra}[1]{\langle\,#1\,|}
\newcommand{\ket}[1]{|\,#1\,\rangle}
\newcommand{\braket}[2]{\ensuremath{\langle\, #1 \mid  #2\, \rangle }}
\newcommand{\moy}[1]{\langle\,#1\,\rangle}

\def\eps{\epsilon}

\begin{document}

\begin{flushright}
LPENSL-TH-04/13
\end{flushright}

\vspace{24pt}

\begin{center}
\begin{LARGE}
{\bf Spontaneous staggered polarizations
%\vspace{.5mm}
of the cyclic solid-on-solid model
from algebraic Bethe Ansatz}
\end{LARGE}

\vspace{50pt}

\begin{large}
{\bf D.~Levy-Bencheton}\footnote{damien.levybencheton@ens-lyon.fr}
{\bf and V.~Terras}\footnote{veronique.terras@ens-lyon.fr}
\end{large}

\vspace{.5cm}
 
{Laboratoire de Physique, ENS Lyon \& CNRS UMR 5672,\\
Universit\'e de Lyon, France}
\vspace{2cm}

\today

\end{center}

\vspace{1cm}

\begin{abstract}
%Using the finite-size determinant representation for the form 
We compute the spontaneous staggered polarization of the cyclic SOS model at the thermodynamic limit. We use the determinant representation for finite-size form factors obtained from algebraic Bethe Ansatz.

\end{abstract}

\vspace{1cm}

%%%%%%%%%%%%%%%%%%%%%%%%%%%%%%%%%%%%%%%%%%%%%%%%%
\section{Introduction}
\label{sec-intro}

In a previous paper \cite{LevT13a}, we have initiated the study of form factors and correlation functions of the cyclic solid-on-solid (CSOS) model \cite{KunY88,PeaS89} by means of algebraic Bethe Ansatz (ABA).
In particular, we have obtained determinant representations for the scalar products of Bethe states and the form factors in finite volume.
The aim of this paper is to show how to apply these results to the computation of physical quantities at the thermodynamic limit. As a simple example, we consider here the spontaneous staggered polarizations.

The CSOS model is a two-dimensional square lattice with interactions around faces (IRF).
On each vertex on the lattice, there is a height $s$ which may take $L$ different values, so that heights on adjacent sites differ by $\pm 1$. The difference of heights between two adjacent sites of the lattice is hence described by a variable $\eps=\pm 1$ attached to the corresponding bond. In the cyclic SOS model \cite{KunY88,PeaS89}, heights are periodic in $L$ (i.e. there exists $s_0\in\mathbb{C}$ such that $s-s_0\in\mathbb{Z}/L\mathbb{Z}$).
There are six different allowed configurations around a face, of the type

\begin{pgfpicture}{0cm}{0cm}{2cm}{2cm}

\pgfnodecircle{Node1}[fill]{\pgfxy(2.5,1.5)}{0.05cm}
\pgfnodecircle{Node2}[fill]{\pgfxy(3.5,1.5)}{0.05cm}
\pgfnodecircle{Node4}[fill]{\pgfxy(2.5,0.5)}{0.05cm}
\pgfnodecircle{Node3}[fill]{\pgfxy(3.5,0.5)}{0.05cm}
\pgfnodeconnline{Node1}{Node2}
\pgfnodeconnline{Node2}{Node3}
\pgfnodeconnline{Node3}{Node4}
\pgfnodeconnline{Node1}{Node4}
% \pgfputat{\pgfxy(0.35,1)}{\pgfbox[left,center]{
% ${R}(u_i-\xi_j; s)^{\epsilon_i,\epsilon_j}_{\epsilon'_i,\epsilon'_j}=$}}
% \pgfline{\pgfxy(3.5,0.5)}{\pgfxy(3.5,1.5)}
% \pgfline{\pgfxy(2.5,0.5)}{\pgfxy(2.5,1.5)}
% \pgfline{\pgfxy(2.5,0.5)}{\pgfxy(3.5,0.5)}
% \pgfline{\pgfxy(2.5,1.5)}{\pgfxy(3.5,1.5)}
 
% \pgfsetendarrow{\pgfarrowto}
% \pgfsetdash{{3pt}{3pt}}{0pt}
% \pgfline{\pgfxy(6.7,1)}{\pgfxy(5.3,1)}%fleche horizontale
% \pgfline{\pgfxy(6,1.7)}{\pgfxy(6,0.3)}%fleche verticale
% \pgfstroke

 \pgfputat{\pgfxy(2.4,1.6)}{\pgfbox[right,center]{$s$}}
 \pgfputat{\pgfxy(3.6,1.6)}{\pgfbox[left,center]{$s+\epsilon'_i$}}
% \pgfputat{\pgfxy(6,0)}{\pgfbox[center,center]{$u_i$}}
 \pgfputat{\pgfxy(2.45,0.1)}{\pgfbox[right,bottom]{$s+\epsilon_j$}}
 \pgfputat{\pgfxy(3.6,0.2)}{\pgfbox[left,bottom]{$s+\epsilon_i+\epsilon_j$}}
 \pgfputat{\pgfxy(3.6,-0.2)}{\pgfbox[left,bottom]{$=s+\epsilon'_i+\epsilon'_j$}}
% \pgfputat{\pgfxy(5,1)}{\pgfbox[center,center]{$\xi_j$}}

\pgfputat{\pgfxy(3,1.6)}{\pgfbox[center,bottom]{$\epsilon'_i$}}
\pgfputat{\pgfxy(3,0.4)}{\pgfbox[center,top]{$\epsilon_i$}}
\pgfputat{\pgfxy(2.4,0.95)}{\pgfbox[right,center]{$\epsilon_j$}}
\pgfputat{\pgfxy(3.6,0.95)}{\pgfbox[left,center]{$\epsilon'_j$}}

\pgfputat{\pgfxy(9.06,1.2)}{\pgfbox[center,center]{with $\epsilon_i,\epsilon'_i,\epsilon_j,\epsilon'_j\in\{+1,-1\}$}}

\pgfputat{\pgfxy(9,0.6)}{\pgfbox[center,center]{such that $\epsilon_i+\epsilon_j=\epsilon'_i+\epsilon'_j$.}}
 
%\pgfputat{\pgfxy(9,1)}{\pgfbox[left,center]{
% $\equiv{W}\binom{s\qquad s+\epsilon'_i}{s+\epsilon_j\ s+\eps_i+\epsilon_j}.$}} 
\end{pgfpicture}

\smallskip
\noindent
To these configurations correspond six statistical weights $W\binom{s\quad\ s+\epsilon'_i}{s+\epsilon_j\ s+\epsilon_i+\epsilon_j } $ which can be parameterized in terms of elliptic theta functions.
They depend on a parameter $\eta$ (crossing parameter) which, in the cyclic case, is a rational number: $\eta=r/L$ with $r,L$ coprime integers. Hence, the statistical weights of the CSOS model are $L$-periodic functions of the height $s$.

It has been shown in \cite{PeaS89,PeaB90} that the transfer matrix of the model with periodic boundary conditions possesses $2(L-r)$ largest (in magnitude) eigenvalues which are asymptotically degenerate in the thermodynamic limit, corresponding to $2(L-r)$ coexisting phases.
These largest eigenvalues are separated from the other ones by a gap which remains finite in the thermodynamic limit.
%In this limit, there exist 
%
The corresponding $2(L-r)$ (quasi-)ground states can be constructed by means of Bethe Ansatz.
In this framework, the spontaneous staggered polarizations can be computed, as in the case of the $F$-model \cite{Bax73}, from the normalized matrix elements, between two of the corresponding Bethe eigenstates, of the $\sigma^z_m$ Pauli spin operator acting on the spin (or variable $\epsilon_m=\pm 1$) on a given bond $m$ of the lattice.
%The obtained normalized Bethe eigenstates $\ket{\psi_g^{(\mathsf{k},\ell)}}$ ($\mathsf{k}\in\mathbb{Z}/2\mathbb{Z}$, $\ell\in\mathbb{Z}/(L-r)\mathbb{Z}$) are not polarized: the mean values
%
%\begin{equation}
%$  \bra{\psi_g^{(\mathsf{k},\ell)}} \sigma_m^z \ket{\psi_g^{(\mathsf{k},\ell)}}, $
%\end{equation}
%
%measuring the polarization on a given bond $m$ of the lattice, vanish in the thermodynamic limit.

The computation of spontaneous staggered polarizations in the CSOS model has already been considered in \cite{DatJKM90} in the case $\eta=1/L$. The derivation of \cite{DatJKM90} uses the representation of Bethe vectors in the framework of coordinate Bethe Ansatz. However, since no compact and convenient representation was known at that time for the scalar products (and form factors) between such Bethe eigenstates, the derivation of \cite{DatJKM90} actually relies on two conjectural mathematical identities for elliptic functions.
The use of the determinant representations obtained in \cite{LevT13a} from algebraic Bethe Ansatz enables us instead to easily compute (for any rational value of the parameter $\eta$ of the model) the thermodynamic limit of the form factors contributing to spontaneous staggered polarizations.
By this method, we are also able to control the finite-size corrections to the result in the same way as in \cite{IzeKMT99}.

This paper is organized as follows. 
In Section~\ref{sec-ABA}, we briefly recall the ABA solution of the CSOS model and the result of \cite{LevT13a} concerning the finite-size determinant representation for the normalized matrix element of the local operator $\sigma_m^z$ between two Bethe eigenstates.
In Section~\ref{sec-gs-th}, we characterize the $2(L-r)$ (quasi-)ground states from the study of their Bethe equations, and discuss the process of taking the thermodynamic limit and of controlling the finite-size corrections.
In Section~\ref{sec-ff-therm}, we apply this process to the form-factor representation of Section~\ref{sec-ABA} in the case where the two Bethe states belong to the previously described set of ground states. We obtain a Fredholm determinant representation that can be explicitly computed, leading, in Section~\ref{sec-pol}, to explicit expressions for the spontaneous polarizations of the model.

%%%%%%%%%%%%%%%%%%%%%%%%%%%%%%%%%%%%%%%%%%%%%%%%%
\section{The form factor in the ABA framework}
\label{sec-ABA}

Let us consider a two-dimensional square lattice of size $N\times N$ ($N$ even), with periodic boundary conditions.
The statistical weights  $W\binom{s\quad\ s+\epsilon'_i}{s+\epsilon_j\ s+\epsilon_i+\epsilon_j } $, $\epsilon_i,\epsilon'_i,\epsilon_j,\epsilon'_j\in\{+1,-1\}$, corresponding to the six allowed configurations around a face of the SOS model, can be seen as the six non-zero elements ${R}(u_i-\xi_j; s)^{\epsilon_i,\epsilon_j}_{\epsilon'_i,\epsilon'_j}$ of the following $R$-matrix:
\begin{equation}\label{R-mat}
  R(u_i-\xi_j;s)=
  \begin{pmatrix} 1 & 0 & 0 & 0 \\
                              0 & b(u_i-\xi_j;s) & c(u_i-\xi_j;s) & 0 \\
                              0 & {c}(u_i-\xi_j;-s) & {b}(u_i-\xi_j;-s) & 0 \\
                              0 & 0 & 0 & 1 
  \end{pmatrix}
  \in\mathrm{End}(\mathbb{C}^2\otimes \mathbb{C}^2).
\end{equation}
Here $u_i$ (respectively $\xi_j$) is an inhomogeneity parameter attached to the column $i$ (resp. row $j$) of cells of the lattice. The functions $b(u;s)$ and $c(u;s)$ are given as
\begin{equation}
 b(u;s)=\frac{[s+1] \,  [u]}{[s] \, [u+1]},           \qquad   c(u;s)=\frac{[s+u] \,  [1]}{[s] \, [u+1]} , \label{bc1}
% &\bar{b}(u;s)=\frac{[s-1] \,  [u]}{[s] \, [u+1]}=b(u;-s) , \quad  &  
% &\bar{c}(u;s)=\frac{[s-u] \,  [1]}{[s] \, [u+1]}=c(u;-s) . \label{bc2}
 \qquad \text{with} \quad [u]=\theta_1(\eta u;\tau),
\end{equation} 
where $\theta_1$ denotes the usual theta function \eqref{theta1} with quasi-periods 1 and $\tau$ ($\Im\tau>0$).
The height $s$ in \eqref{R-mat} is called dynamical parameter. 
The $R$-matrix \eqref{R-mat} with dynamical parameter $s$  satisfies the dynamical Yang-Baxter equation \cite{GerN84,Fel95}, and the corresponding SOS model is also sometimes called dynamical six-vertex model. In the cyclic case that we consider here, the parameter $\eta$ of the model is chosen to be rational: $\eta=r/L$, $r$ and $L$ being relatively prime integers.  Hence, the statistical weights elements of the $R$-matrix are $L$-periodic functions of $s$.

\begin{rem}
Our parameter $\eta s_0$, where $s_0$ is a global shift of the dynamical parameter (such that $s-s_0$ is an integer) introduced so as to avoid the singularities in \eqref{bc1}, is related to the phase angle $\mu=w_0/\pi$ of the physical model introduced in \cite{PeaS89,PeaB90} by a shift of $\tau/2$ (later on, we will for simplicity take $s_0=\frac{\tau}{2\eta}$, i.e. $\mu=0$, in our result). Also, the statistical weights of  \cite{PeaS89,PeaB90} correspond to a diagonal dynamical gauge transformation of the $R$-matrix \eqref{R-mat} which leaves the local height probabilities invariant. 
\end{rem}

Eigenstates of the dynamical transfer matrix constructed from the $R$-matrix \eqref{R-mat}
%, which corresponds to the product of all statistical weights along a column of elementary cells of the lattice, 
can be obtained by means of algebraic Bethe Ansatz (see \cite{FelV96a,FelV96b}).
In this framework, the space of states of the model corresponds to the space of functions $\mathrm{Fun}(\mathcal{H}[0])$ of the dynamical parameter $s\in s_0+\mathbb{Z}/L\mathbb{Z}$ with values in the zero-weight space $\mathcal{H}[0]=\{\,\ket{\!\mathbf{v} \! }\in\mathcal{H}\mid  (\sum_{j=1}^N\sigma_j^z)\,\ket{\! \mathbf{v} \! }=0\}$, with $\mathcal{H}\sim (\mathbb{C}^2)^{\otimes N}$ (see \cite{LevT13a} for more details).
More precisely, from a product of $R$-matrices along a column of elementary cells of the lattice, one can construct the monodromy matrix, which is a $2\times 2$ matrix of operators 
\begin{equation}\label{mon-op}
   \widehat{T}(u)= \begin{pmatrix} \widehat{A}(u) & \widehat{B}(u) \\
                                    \widehat{C}(u) & \widehat{D}(u) \end{pmatrix}             
\end{equation}
with $\widehat{A}(u), \widehat{B}(u), \widehat{C}(u), \widehat{D}(u)\in \mathrm{End}(\mathrm{Fun}\mathcal{H})$.
Common eigenstates to transfer matrices  $\widehat{t}(u)=\widehat{A}(u)+\widehat{D}(u)$ can be constructed in $\mathrm{Fun}(\mathcal{H}[0])$ as
\begin{equation}\label{state}
  \ket{ \{v\}, \omega }: 
 s \mapsto %\ket{ \{ v_1,\ldots,v_n\};s} = 
 \omega^s \prod_{j=1}^{n} \frac{[1]}{[s-j]} \, B(v_1;s) B(v_2;s-1)\ldots B(v_n;s-n+1) \ket{0}.
\end{equation}
Here $n$ is such that $N=2n+\aleph L$ for some integer $\aleph$, $\omega$ is such that $(-1)^{rn} \omega^L=1$, the reference state $\ket{0}\in\mathcal{H}$ is chosen as $\ket{0}=\otimes_{i=1}^N\binom{1}{0}$, and $\{v\}\equiv\{v_1,\ldots,v_n\}$ (with $\eta v_i\neq\eta v_j\mod\mathbb{Z}+\tau\mathbb{Z}$) is a solution of the system of Bethe equations
\begin{equation}\label{Bethe}
    a(v_j)  \prod_{l\ne j} \frac{[v_l-v_j+1]}{[v_l-v_j]} 
    =  (-1)^{r\aleph} \omega^{-2} \; d(v_j)  \prod_{l\ne j} \frac{[v_j-v_l+1]}{[v_j-v_l]} ,
    \quad j=1,\ldots n,
\end{equation}
with
\begin{equation}\label{a-d}
  a(u)=1, \qquad d(u)=\prod_{j=1}^N \frac{[u-\xi_j]}{[u-\xi_j+1]}.
\end{equation}
Similarly, eigenstates of the transfer matrix in the dual space are constructed as
\begin{equation}\label{dual}
   \bra{\{ v\} ,\omega}:
   s\mapsto \bra{0}C(v_n;s-n) \ldots  C(v_2;s-2) C(v_1;s-1) \,
     {\omega}^{- s}  \prod_{j=0}^{n-1} \frac{[s+j]}{[1]},
\end{equation}
with  $\bra{0}=\ket{0}^\dagger$ and $\{v\}$ solution of \eqref{Bethe}.
Then
\begin{equation}\label{act-transfer}
   \widehat{t}(u)\, \ket{\{ v\},\omega }= \tau (u; \{ v\},\omega )\, \ket{\{ v\},\omega },
   \qquad
   \bra{ \{v\}, \omega }\,\widehat{t} (u) = \tau(u; \{v\},\omega )\,  \bra{ \{v\} ,\omega },
\end{equation}
with
\begin{equation}
   \tau (u; \{v\},\omega)
    = \omega \; a(u) \prod_{l=1}^n \frac{[v_l-u+1]}{[v_l-u]}
        +  (-1)^{r\aleph}  \omega^{-1} \; d(u) \prod_{l=1}^n \frac{[u-v_l+1]}{[u-v_l]}.
    \label{tau}
\end{equation}
%

%In this framework, the spontaneous polarizations of the model an be computed from the nprmalized matrix elements of the local spin operator $\sigma_m^z$ between two 
From now on, we restrict our study to Bethe states in the sector $n=N/2$, which is enough for our purpose.
For two such Bethe eigenstates, associated to two different solutions $\{u\},\omega_u$ and $\{v\},\omega_v$ of the system of Bethe equations \eqref{Bethe}, we want to compute the following renormalized form factor:
\begin{align}\label{magnetization}
   \mathbf{s}_m^z(\{u\},\omega_u;\{v\},\omega_v) 
   &=  \frac{\bra{\{u\},\omega_u}\, \sigma_m^z \,\ket{\{v\},\omega_v} }{ \big( \braket{ \{u\},\omega_u}{ \{u\},\omega_u } \braket{ \{v\},\omega_v}{ \{v\},\omega_v } \big)^{1/2}} 
   \\
   &=\frac{\bra{\{u\},\omega_u}\, \sigma_m^z \,\ket{\{v\},\omega_v} }{  \braket{ \{v\},\omega_v}{ \{v\},\omega_v } }
      \cdot
    \left(\frac{ \braket{ \{v\},\omega_v}{ \{v\},\omega_v } }{\braket{ \{u\},\omega_u}{ \{u\},\omega_u } } \right)^{1/2}.\label{magn-bis}
\end{align}
This quantity can be rewritten in terms of a ratio of determinants by means of the representations obtained in our previous paper \cite{LevT13a}.
We recall that the matrix element of the operator $\sigma_m^z$ between two different Bethe eigenstates $\bra{\{u\},\omega_u}$ \eqref{dual} and $\ket{\{v\},\omega_v}$ \eqref{state} in the sector $N=2n$ 
% with an arbitrary state of Bethe type (i.e. a state of the form \eqref{state} with $v_1, \ldots, v_n$ arbitrary parameters 
is given as
\begin{multline}
   \bra{\{u\},\omega_u }\, \sigma_m^{z} \, \ket{\{v\},\omega_v }
   = \Bigg\{ \prod_{k=1}^{m-1} \frac{\tau(\xi_k;\{ u \},\omega_u ) }{ \tau(\xi_k;\{ v \},\omega_v )}\Bigg\}  
          \Bigg\{ \frac{1}{L}\sum_{s\in s_0+\mathbb{Z}/L\mathbb{Z} } \hspace{-1mm}
          \omega_u^{-s}\omega_v^{s} \frac{[\gamma+s ]}{[s]} \Bigg\}
         \\
   \times
         \frac{\prod_{t=1}^n d(u_t)}{\prod_{k<l}[u_k-u_l][v_l-v_k]}   
         \det_n \big[  \Omega_\gamma(\{u\},\omega_u;\{v\},\omega_v)
                              -2\mathcal{P}_\gamma(\{u\},\omega_u;\{v\},\omega_v|\xi_m)\big],
         \label{ff-sigmaz}
\end{multline}
with $\gamma=\sum_{j=1}^n(v_j-u_j)$. Here $\Omega_\gamma$ is the matrix appearing in the determinant representation for the scalar product (see Theorem 3.1 of \cite{LevT13a}), with matrix elements%
\begin{multline}\label{def-Omega}
   \hspace{-2mm}[ \Omega_\gamma (\{u \},\omega_u; \{ v \},\omega_v  ) \big]_{jk} 
   = \frac{1}{[\gamma]}\Bigg\{\! \frac{[u_{j}-v_{k}+\gamma]}{[u_{j}-v_{k}]}
                  -\frac{\omega_v}{\omega_u}\frac{[u_{j}-v_{k}+\gamma+1]}{[u_{j}-v_{k}+1]} \! \Bigg\} 
       a(v_{k}) \! \prod_{t=1}^{n} [u_{t}-v_{k}+1]
       \\
     + \frac{1}{[\gamma]}
     \Bigg\{\! \frac{[u_{j}-v_{k}+\gamma]}{[u_{j}-v_{k}]}
                  -\frac{\omega_u}{\omega_v}\frac{[u_{j} -v_{k}+\gamma-1]}{[u_{j}-v_{k}-1]} \!  \Bigg\}    
        \omega_u^{-2} d(v_{k}) \! \prod_{t=1}^{n} [u_{t}-v_{k}-1],
\end{multline}
whereas $ \mathcal{P}_\gamma$ is a rank 1 matrix defined as
\begin{multline}\label{mat-P}
    \big[ \mathcal{P}_\gamma(\{u\},\omega_u;\{v\},\omega_v|\xi_m) \big]_{jk} 
    =   \frac{1}{[\gamma]} \bigg\{ 
          \frac{[u_{j} - \xi_m + \gamma]}{[u_{j} - \xi_m]} 
       - \frac{\omega_v}{\omega_u} \frac{[u_{j} - \xi_m + \gamma + 1]}{[u_{j} - \xi_m + 1 ]}  \bigg\} 
         \\
        \times
        a(v_{k}) \prod_{t=1}^n \bigg\{ [v_{t} - v_{k}+1] \frac{[u_{t} - \xi_m +1 ]}{[v_{t} - \xi_m + 1]}  \bigg\}  .
\end{multline}
%
%Note that we have particularized the expressions \eqref{def-Omega} and \eqref{mat-P} to  the case $N=2n$, which is the case under interest for (see Section~\ref{sec-gs-th}).
We also recall the determinant representation for the `square of the norm' of a Bethe eigenstate:
% (still in the case $N=2n$):
%
\begin{equation}
  \moy{\{u\},\omega_u \, | \, \{u\},\omega_u}
     =        
     \frac{ \prod_{t=1}^{n}  a(u_{t})  d(u_{t}) \,
                    \prod_{j,k=1}^n [u_{j}-u_{k}+1] }
             { (-[0]')^n\,\prod_{j\not= k} [u_{j}-u_{k}]  }   
     \det_n \big[ \Phi (\{u\} ) \big]   ,\label{gaudin}
\end{equation}  
with
\begin{multline}\label{mat-Phi}
   \big[ \Phi (\{u\} ) \big]_{jk}
   = \delta_{jk}\Bigg\{ \log'\frac{a}{d}(u_{j})
                                    +\sum_{t=1}^n \left(\frac{[u_j-u_t-1]'}{[u_j-u_t-1]}-\frac{[u_j-u_t+1]'}{[u_j-u_t+1]}\right)\Bigg\}\\
      -  \left(\frac{[u_j-u_k-1]'}{[u_j-u_k-1]}-\frac{[u_j-u_k+1]'}{[u_j-u_k+1]}\right).
\end{multline}   

Factorizing the quantity $a(v_j)\prod_{t=1}^n [v_t-v_j+1] = - \omega_v^{-2} d(v_j)\prod_{t=1}^n [v_t-v_j-1]$ out of each column of the determinant in \eqref{ff-sigmaz}, we can rewrite the first ratio in \eqref{magn-bis} as:
\begin{multline}\label{ratio1}
 \frac{\bra{\{u\},\omega_u}\, \sigma_m^z \,\ket{\{v\},\omega_v} }{  \braket{ \{v\},\omega_v}{ \{v\},\omega_v } }
  =
   \left\{ \prod_{k=1}^{m-1} \frac{\tau(\xi_k;\{u\},\omega_u)}{\tau(\xi_k;\{v\},\omega_v)} \right\}
 \Bigg\{ \frac{1}{L}\sum_{s\in s_0+\mathbb{Z}/L\mathbb{Z} } \left(\frac{\omega_v}{\omega_u}\right)^{\! s} \frac{[\gamma+s]}{[s]}\Bigg\}
 \\
 \times (-[0]')^n\prod_{k=1}^n\frac{d(u_k)}{d(v_k)}\prod_{k<l}\frac{[v_k-v_l]}{[u_k-u_l]} 
 \frac{\det_n[H(\{u\},\{v\})-2Q(\{u\},\{v\})]}{\det_n [\Phi(\{v\})]},
\end{multline}
where $\gamma=\sum_{t=1}^n(v_t-u_t)$, $\Phi$ is given by \eqref{mat-Phi}, and
\begin{multline}\label{mat-H}
   [H(\{u\},\{v\})]_{ij}=\frac{1}{[\gamma]}\Bigg\{\! \frac{[u_{i}-v_{j}+\gamma]}{[u_{i}-v_{j}]}
                  -\frac{\omega_v}{\omega_u}\frac{[u_{i}-v_{j}+\gamma+1]}{[u_{i}-v_{j}+1]} \! \Bigg\} 
       \prod_{t=1}^{n} \frac{ [u_{t}-v_{j}+1]}{[v_t-v_j+1]}
       \\
     - \frac{1}{[\gamma]}
     \Bigg\{\! \frac{[u_{i}-v_{j}+\gamma]}{[u_{i}-v_{j}]}
                  -\frac{\omega_u}{\omega_v}\frac{[u_{i} -v_{j}+\gamma-1]}{[u_{i}-v_{j}-1]} \!  \Bigg\}    
        \frac{\omega_v^2}{\omega_u^2}  \prod_{t=1}^{n}\frac{ [u_{t}-v_{j}-1]}{[v_t-v_j-1]},
\end{multline}
\begin{equation}
   [Q(\{u\},\{v\})]_{ij}=\frac{1}{[\gamma]} \bigg\{ 
          \frac{[u_{i} - \xi_m + \gamma]}{[u_{i} - \xi_m]} 
       - \frac{\omega_v}{\omega_u} \frac{[u_{i} - \xi_m + \gamma + 1]}{[u_{i} - \xi_m + 1 ]}  \bigg\} 
         \prod_{t=1}^n \frac{[u_{t} - \xi_m +1 ]}{[v_{t} - \xi_m + 1]} .
\end{equation}
In its turn, from \eqref{gaudin}, the second ratio is given by
\begin{equation}\label{ratio2}
  \frac{ \braket{ \{v\},\omega_v}{ \{v\},\omega_v } }{\braket{ \{u\},\omega_u}{ \{u\},\omega_u } }
  =
  \prod_{k=1}^n\! \frac{a(v_k) d(v_k)}{a(u_k) d(u_k) }
  \prod_{j,k=1}^n\! \frac{[v_j-v_k+1]}{[u_j-u_k+1]}\prod_{j\not= k}\! \frac{[u_j-u_k]}{[v_j-v_k]}
  \frac{\det_n [\Phi(\{v\})]}{\det_n [\Phi(\{u\})]}.
\end{equation}
These representations are the starting point for our study of the spontaneous staggered polarizations of the model, which can be obtained from the thermodynamic limit of the quantity \eqref{magnetization}
in the case when $\ket{\{u\},\omega_u}$ and $\ket{\{v\},\omega_v}$ correspond to two different (quasi-)ground states of the homogeneous model.

\begin{rem}\label{rem-gamma=0}
The case of the mean value of the operator $\sigma_m^z$ in the same Bethe state should be treated separately since the proper limit has to be taken into the determinant of \eqref{ff-sigmaz} in the same way as in \eqref{gaudin}-\eqref{mat-Phi}.
We will in that case use the formula
\begin{equation}\label{mean-sgz}
 \mathbf{s}_m^z(\{u\},\omega_u;\{u\},\omega_u) 
%   \frac{\bra{\{u\},\omega_u}\, \sigma_m^z \,\ket{\{u\},\omega_u} }{  \braket{ \{u\},\omega_u}{ \{u\},\omega_u } }
  =
\frac{\det_n[\Phi(\{u\})+2Q^{(+)}(\{u\})]}{\det_n [\Phi(\{u\})]},
\end{equation}
with $\Phi$ given by \eqref{mat-Phi} and
\begin{equation}
  \big[Q^{(+)}(\{u\}) \big]_{jk}=\frac{[u_j-\xi_m]'}{[u_j-\xi_m]}-\frac{[u_j-\xi_m+1]'}{[u_j-\xi_m+1]}.
\end{equation}
%
%This should be also the case when some of the $v_j$ coincide with some the $u_j$.
One should also pay special care to possible other cases for which $\gamma=0$ or when some of the $v_j$ coincide with some the $u_j$.
In particular, when $L$ is even, there is a little subtlety that what not mentioned in our previous article \cite{LevT13a}: in that case, with each solution $\{u\},\omega^2$ of the Bethe equations \eqref{Bethe}, one can associate two different Bethe eigenstates $\ket{\{u\},\omega}$ (with $\omega=e^{i\pi(rn+2\ell)/L}$ for some integer $\ell$) and $\ket{\{u\},-\omega}$ ($-\omega=e^{i\pi(rn+2\ell')/{L}}$ with $\ell'=\ell-L/2$) with opposite eigenvalue.
% each Bethe eigenstate $\ket{\{u\},\omega}$, with $\{u\},\omega=e^{i\pi(rn+2\ell)/L}$, $\ell\in\{0,1,\ldots,L-1\}$, solution of the Bethe equations \eqref{Bethe}, one can associate a different Bethe eigenstate $\ket{\{u\},-\omega}$with opposite eigenvalue ($-\omega=e^{i\pi(rn+2\ell')/{L}}$ with $\ell'=\ell+L/2$).
The corresponding normalized form factor can be explicitly represented as
\begin{equation}
 \mathbf{s}_m^z(\{u\},\omega_u;\{u\},-\omega_u) 
    =  \frac{(-1)^{m-1} }{L}\sum_{s\in s_0+\mathbb{Z}/L\mathbb{Z} } \hspace{-1mm}
          e^{-i\pi s} \
     \frac{\det_n \big[ \Phi^{(-)} (\{u\} )+2 Q^{(-)}(s) \big]}{\det_n\big[\Phi(\{u\})\big]},
         \label{ff-sigmaz-bis}
\end{equation}
with $\Phi$ given by \eqref{mat-Phi} and
\begin{align}
 &\big[ \Phi^{(-)} (\{u\} ) \big]_{jk}
   = \delta_{jk}\Bigg\{ \log'\frac{a}{d}(u_{j})
                                    +\sum_{t=1}^n \left(\frac{[u_j-u_t-1]'}{[u_j-u_t-1]}-\frac{[u_j-u_t+1]'}{[u_j-u_t+1]}\right)\Bigg\} \nonumber\\
  &\hspace{7cm}    +  \left(\frac{[u_j-u_k-1]'}{[u_j-u_k-1]}-\frac{[u_j-u_k+1]'}{[u_j-u_k+1]}\right),
  \\
  &\big[Q^{(-)}(s) \big]_{jk}=2\frac{[s ]'}{[s]}.
\end{align}
\end{rem}

%%%%%%%%%%%%%%%%%%%%%%%%%%%%%%
\section{The degenerate ground states in the thermodynamic limit}
\label{sec-gs-th}

According to \cite{PeaS89,PeaB90}, the ground state of the CSOS model is degenerate at the thermodynamic limit: there are $2(L-r)$ (quasi-)ground states in the sector $n=N/2$, which were identified in \cite{PeaS89,PeaB90} in the low-temperature limit.

To characterize these states, it is convenient, by means of Jacobi's imaginary transformation
\begin{equation}\label{jacobi-tr}
   [u] = -i (-i\tau)^{-1/2} e^{i\pi\eta\tilde\eta u^2} \theta_1(\tilde\eta u;\tilde\tau),
   \qquad
   \tilde\tau=-\frac{1}{\tau},
   \quad
   \tilde\eta=-\frac{\eta}{\tau},
\end{equation}
to rewrite the Bethe equations \eqref{Bethe} in terms of theta functions with imaginary quasi-period $\tilde\tau$.
In the sector $n=N/2$ and at the homogeneous limit $\xi_k=1/2$, $k=1,\ldots,N$, it gives
\begin{equation}\label{Bethe-bis}
   \omega^2 e^{4i\pi\eta \sum_{l=1}^n z_l  }\,
  %e^{2i\pi\eta \{ (N-2n) z_j+\tilde{\eta}(\frac{N}2-\sum_{k=1}^N\xi_k)+2\sum_{l=1}^n z_l \} } \,
  \frac{\theta_1^N(z_j+\tilde\eta/2)}{\theta_1^N(z_j-\tilde\eta/2)}
  \prod_{l=1}^n \frac{\theta_1(z_l-z_j+\tilde\eta)}{\theta_1(z_l-z_j-\tilde\eta)}
  = -1, %(-1)^{r\aleph+1}.
\end{equation}
where we have set $z_j=\tilde\eta v_j$, $j=1,\ldots, n$. Here and in the following, unless explicitly specified, the considered theta functions are of imaginary quasi-period $\tilde\tau$, i.e. $\theta_1(z)\equiv \theta_1(z;\tilde\tau)$.

These Bethe equations can be rewritten in the logarithmic form as
\begin{equation}\label{Bethe-log}
  N p_0(z_j)-\sum_{l=1}^n \vartheta(z_j-z_l)=2\pi \Big(n_j -\frac{n+1}{2}+\beta+2\eta\sum_{l=1}^nz_l\Big),\quad\  j=1,\ldots , n,
\end{equation}
where $n_j$ are integers, $p_0$ and $\vartheta$ are the bare momentum and bare phase
\begin{equation}\label{bare-mom-ph}
  p_0(z)=i\log\frac{\theta_1(\tilde\eta/2+z)}{\theta_1(\tilde\eta/2-z)},
  \qquad
  \vartheta(z)=i\log\frac{\theta_1(\tilde\eta+z)}{\theta_1(\tilde\eta-z)},
\end{equation}
and $\omega=e^{i\pi \beta}$.
The degenerate ground states identified in \cite{PeaB90} correspond to real solutions of \eqref{Bethe-log} such that $n_{j+1}-n_j =  1$,  $j =1,\ldots, n-1$.
In the thermodynamic limit $N\to\infty$ (with $n=N/2$), the distribution of the Bethe roots corresponding to such states tends to a  positive density $\rho(z)$ on the interval $[-1/2,1/2]$, solution of the following integral equation:
\begin{equation}\label{liebeq}
   \rho(z)+\int_{-1/2}^{1/2} K(z-w)\, \rho(w)\, dw =\frac{p'_0(z)}{2\pi},
\end{equation} 
with
\begin{align}
  &p_0'(z)=i \left\{ \frac{\theta_1'(z+\tilde\eta/2)}{\theta_1(z+\tilde\eta/2)}   -\frac{\theta_1'(z-\tilde\eta/2)}{\theta_1(z-\tilde\eta/2)}\right\},\\
  &K(z)=\frac{1}{2\pi}\vartheta'(z)
         = \frac{i}{2\pi} \left\{ \frac{\theta_1'(z+\tilde\eta)}{\theta_1(z+\tilde\eta)}   -\frac{\theta_1'(z-\tilde\eta)}{\theta_1(z-\tilde\eta)}\right\}. \label{K}
\end{align}
The solution of the integral equation \eqref{liebeq} can easily be computed by means of the Fourier transform. In the domain $0<\eta<1/2$, i.e. $0<-i\tilde\eta<-\frac{i}{2}\tilde\tau$, we have
\begin{align}
   &p_m'=\int_{-1/2}^{1/2} p'_0(z)\, e^{-2\pi i m z}\, dz
             = \begin{cases}
                 2\pi & \text{for } m=0,\\
                 2\pi \tilde{q}^{\frac{|m|}2}\frac{1- \tilde{p}^{|m|} \tilde{q}^{-|m|}}{1- \tilde{p}^{|m|}}  & \text{otherwise,}
  \end{cases}\\
  & k_m = \int_{-1/2}^{1/2} K(z)\, e^{-2\pi i m z}\, dz
              = \begin{cases}
                 1 & \text{for } m=0,\\
                  \tilde{q}^{|m|}\frac{1- \tilde{p}^{|m|} \tilde{q}^{-2|m|}}{1- \tilde{p}^{|m|}}  & \text{otherwise,}
                 \end{cases} \label{km}
\end{align}
such that
\begin{equation}\label{rho}
   \rho(z)=\sum_{m=-\infty}^\infty \frac{e^{2\pi i m z}}{2\cosh(i\pi m\tilde\eta)}
%   =\frac{1}{2}\prod_{m=1}^{\infty}\frac{(1- \tilde{q}^m)^2\,(1+2 \tilde{q}^{m-\frac12}\cos 2\pi z + \tilde{q}^{2m-1})}{(1+ \tilde{q}^m)^2\,(1-2 \tilde{q}^{m-\frac12}\cos 2\pi z+ \tilde{q}^{2m-1})},
  = \frac{1}{2}\prod_{m=1}^{\infty}\frac{(1- \tilde{q}^m)^2}{(1+ \tilde{q}^m)^2}\,
     \frac{\theta_3(z;\tilde\eta)}{\theta_4(z;\tilde\eta)},
\end{equation}
in which we have set $ \tilde{q}=e^{2\pi i \tilde\eta}$, $ \tilde{p}=e^{2\pi i \tilde\tau}$.

We now want to study more precisely how the Bethe roots of one of these ground states behave with respect to finite size corrections.
Let us introduce, for a given (quasi-)ground state parametrized by the set of roots $\{x_j\}_{1\leq j \leq n}$ solution to \eqref{Bethe-log} with $\omega_x=e^{i\pi\beta_x}$ and a given shift of integers $\mathsf{k}_x = n_j-j$ (which does not depend on $j$), the following counting function:
\begin{equation}\label{counting}
  \widehat\xi_{x}(z)=\frac{1}{\pi}p_0(z)-\frac{1}{\pi N}\sum_{l=1}^n\vartheta(z-x_l)
                                 +\frac{1}{n}\left(\frac{n+1}{2}-\mathsf{k}_x-\beta_x-2\eta\sum_{l=1}^n x_l \right).
\end{equation}
This function is such that $\widehat\xi_{x}(x_j)=j/n$, $j=1,\ldots, n$.
Moreover, since its derivative,
\begin{equation}\label{der-counting}
 \widehat\xi'_{x}(z)=\frac{1}{\pi}p'_0(z)-\frac{2}{ N}\sum_{l=1}^n K(z-x_l),
\end{equation}
tends to $2\rho(z)$ (which is positive) in the thermodynamic limit, $\widehat\xi_{x}$  is an increasing, and hence invertible function, at least for $N$ large enough.
Hence one can show the following result:
%More generally, one can show the following result, the demonstration being completely analogous to the one presented in Appendix~B of \cite{IzeKMT99}:
%
\begin{prop}\label{prop-sum-int}
Let $f$ be a $\mathcal{C}^\infty$ 1-periodic function on $\mathbb{R}$. Then, the sum of all the values $f(x_j)$, where the set of spectral parameters $\{x_j\}_{1\leq j \leq n}$ parametrizes one of the degenerate ground states solution to \eqref{Bethe-log}, can be replaced by an integral in the thermodynamic limit according to the following rule:
\begin{equation}\label{sum-int}
\frac{1}{N} \sum_{j=1}^n f(x_j) = \int_{-1/2}^{1/2}  f(z)\, \rho(z) \, dz+ O(N^{-\infty}).
\end{equation}
Similarly, if $g$ is a $\mathcal{C}^\infty$ function such that $g'$ is 1-periodic, then
\begin{equation}\label{sum-int2}
\frac{1}{N} \sum_{j=1}^n g(x_j) = \int_{-1/2}^{1/2} g(z)\, \rho(z)\, dz +\frac{c_g}{N}\sum_{j=1}^n x_j+ O(N^{-\infty}),
\end{equation}
where $c_g=\int_{-1/2}^{1/2} g'(z)\, dz = g(1/2)-g(-1/2)$.
\end{prop}
\Proof
The proof of \eqref{sum-int} is similar to the proof of Proposition~3.1 of \cite{IzeKMT99}.
For completeness, we recall its main arguments.
It relies on the fact that one can easily prove, using the Taylor expansion of the 1-periodic function $f$, the analog of \eqref{sum-int} in the case of homogeneously distributed variables:
\begin{equation}\label{sum-int-bis}
\frac{1}{n} \sum_{j=1}^n f\Big(\frac{j}{n}\Big) = \int_{0}^{1}  f(z) \, dz+ O(N^{-\infty}).
\end{equation}
One should then notice that $\widehat{\xi}_{x}$  is a $\mathcal{C}^\infty$ function of real variables such that $\widehat{\xi}_{x}(z+1)=\widehat{\xi}_{x}(z)+1$, so that the function $f \circ \widehat{\xi}_{x}^{-1}$ is also 1-periodic. One can therefore apply \eqref{sum-int-bis} to $f \circ \widehat{\xi}_x^{-1}$ and perform a change of variables in the integral to express the sum over $x_j$ as
\begin{equation}\label{sum-int-ter}
\frac{1}{n} \sum_{j=1}^n f(x_j) =\frac{1}{n} \sum_{j=1}^n f\big(\widehat{\xi}_{x}^{-1}(j/n)\big) 
= \int_{-1/2}^{1/2}  f(z)\, \widehat\xi_{x}'(z) \, dz+ O(N^{-\infty}).
\end{equation}
Finally, applying \eqref{sum-int-ter} to the r.h.s. of \eqref{der-counting}, we obtain that, up to corrections of order $O(N^{-\infty})$, the function $\widehat\xi_{x}' /2$ satisfies the same integral equation \eqref{liebeq} as $\rho$. By unicity of the solution, we have
\begin{equation}\label{der-count-rho}
  \widehat\xi_{x}'(z)=2\rho(z) +O(N^{-\infty}),
\end{equation}
which ends the proof of \eqref{sum-int}.

The identity \eqref{sum-int2} is then a direct corollary of \eqref{sum-int}: if $g'(x)$ is 1-periodic, then $g(x)-c_g x$ is also 1-periodic, and one can apply \eqref{sum-int} to get
\begin{equation}
    \frac{1}{N}\sum_{j=1}^n g(x_j)-\frac{c_g}{N}\sum_{j=1}^n x_j= \int_{-1/2}^{1/2} g(z)\,\rho(z)\, dz -c_g\int_{-1/2}^{1/2} z\,\rho(z)\, dz+ O(M^{-\infty}),
\end{equation}
the last integral being zero by symmetry.
\qed

\begin{rem}
Proposition~\ref{prop-sum-int} can be used to obtain a sum rule for the corresponding ground state roots. Summing all logarithmic Bethe equations \eqref{Bethe-log} for $j=1,\ldots,n$ and using the fact that $\vartheta$ is an odd function, we get
\begin{equation}\label{sum-p0}
 \frac{1}{\pi} \sum_{j=1}^np_0(x_j)= \mathsf{k}_x+\beta_x +2\eta\sum_{j=1}^n x_j .
\end{equation}
Using then \eqref{sum-int2} applied to the l.h.s of \eqref{sum-p0},
\begin{equation}\label{sum-px}
    \sum_{j=1}^np_0(x_j)=2\pi\sum_{j=1}^n x_j + O(N^{-\infty}),
\end{equation}
we obtain that
\begin{equation}
  \sum_{j=1}^n x_j
  = \frac{ \mathsf{k}_x+\beta_x}{ 2(1-\eta)} +O(N^{-\infty})
  = \frac{L\mathsf{k}_x+rn+2\ell_x}{2(L-r)}+O(N^{-\infty}),\label{sum-x}
\end{equation}
in which we have set $\beta_x=\frac{rn+2\ell_x}{L}$.
\end{rem}

\begin{rem}
The counting function \eqref{counting} can be evaluated in the thermodynamic limit as
\begin{equation}\label{count-therm}
   \widehat\xi_{x}(z)=2\int_{0}^z\rho(w)\, dw+\frac{n+1}{N} -\frac2N\sum_{j=1}^n x_j+O(N^{-\infty}).
\end{equation}
This follows from \eqref{der-count-rho} and from the value $\widehat\xi_{x}(0) $ which can be evaluated in the thermodynamic limit by means of  \eqref{sum-x} as well as \eqref{sum-int2} applied to the odd function $\vartheta$:
\begin{align}
  \widehat\xi_{x}(0) 
   &=-\frac{1}{\pi N} \sum_{\ell=1}^n \vartheta(-x_\ell)+\frac1n \bigg( \frac{n+1}2 -\mathsf{k}_x-\beta_x-2\eta\sum_{j=1}^n x_j \bigg) \nonumber\\
   &=\frac1n\bigg( \frac{n+1}2-\sum_{j=1}^n x_j\bigg) + O(N^{-\infty}).
\end{align}
\end{rem}

One can now use these results to compute more precisely the infinitesimal difference of roots $x_{j+1}-x_j$. The latter is given as the value $\widehat\delta_x(x_j)$ of the infinitesimal shift function $\widehat\delta_x$ defined as
\begin{equation}\label{shiftx}
  \widehat\delta_x(z)=\widehat\xi^{-1}_{x}\Big(\widehat\xi_{x}(z)+\frac{1}{n}\Big)-z.
\end{equation}
Rewriting the equation $\widehat\xi_{x}\big(\widehat\delta_x(z)+z\big)=\widehat\xi_{x}(z)+\frac{1}{n}$ using the representation \eqref{count-therm} of $\widehat\xi_{x}$, one gets
\begin{equation}\label{eq-shiftx}
  \int_z^{z+\widehat\delta_x(z)} \hspace{-2mm} \rho(w)\, dw = \frac1N+O(N^{-\infty}).
\end{equation}
Expanding $\rho$ in Taylor series, one therefore obtains a relation which enables one in principle to compute $\widehat\delta_x(z)$ at all order in $N$:
%  
%Taking the difference between the two Bethe equations \eqref{Bethe-log} for the roots $x_{j+1}$ and $x_j$, expanding the differences  of functions $p_0$ and $\vartheta$ as Taylor series and using Proposition~\ref{prop-sum-int}, one obtains that
%
\begin{equation}\label{dev-x}
   N\sum_{k=1}^\infty \, \frac{1}{k !} \rho^{(k-1)}(z) \, \big[\,\widehat\delta_x(z)\big]^k =1 +O(N^{-\infty}).
\end{equation}
%
%where the function $\widehat\delta_x$ is defined as $ \widehat\delta_x(z)=\widehat\xi^{-1}_{\alpha;x}\big(\widehat\xi_{\alpha;x}(z)+\frac{1}{N}\big)-z.$ The relation \eqref{dev-x} hence enables one to obtain the infinitesimal difference of two successive roots $\widehat\delta_x(x_j)=x_{j+1}-x_j$ at all order in $N$.

Let us now consider two different ground states for the system of Bethe equations \eqref{Bethe-log}, parameterized by a solution $\{x_j\}_{j=1,\ldots,n}$, $\omega_x=e^{i\pi\beta_x}$, and a shift of integers $\mathsf{k}_x$ (respectively by $\{y_j\}_{j=1,\ldots,n}$, $\omega_y=e^{i\pi\beta_y}$, and a shift of integers $\mathsf{k}_y$).
%Still following the lines of \cite{IzeKMT99}, we now consider two different (quasi-)ground states parameterized respectively by  the sets of roots $\{x_j\}_{j=1,\ldots,n}$ %with $\omega_x=e^{i\pi\beta_x}$ and $\{y_j\}_{j=1,\ldots,n}$, %with $\omega_y=e^{i\pi\beta_y}$. We 
We want to evaluate, at large $N$, the infinitesimal difference of roots $x_{j}  -  y_{j}$.
To this aim, we define the infinitesimal shift function
\begin{equation}\label{shiftxy}
   \widehat\delta_{x,y}(z)=\widehat\xi_{x}^{-1}\Big(\widehat\xi_{y}(z)\Big)-z,
\end{equation}
which is such that $\widehat\delta_{x,y}(y_j)=x_j-y_j$.
Using again the representation \eqref{count-therm} for $\widehat\xi_{x}$ and $\widehat\xi_{y}$, one obtains for $\widehat\delta_{x,y}$ an equation analog to \eqref{eq-shiftx}:
\begin{equation}\label{eq-shiftxy}
  \int_z^{z+\widehat\delta_{x,y}(z)} \hspace{-2mm} \rho(w)\, dw = \frac1N\sum_{j=1}^n(x_j-y_j)+O(N^{-\infty}),
\end{equation}
which leads to
\begin{equation}  \label{dev-xy}
   N\sum_{k=1}^\infty \, \frac{1}{k !} \rho^{(k-1)}(z) \, \big[\,\widehat\delta_{x,y}(z)\big]^k =\sum_l(x_l-y_l) +O(N^{-\infty}).
\end{equation}
Comparing this equation to \eqref{dev-x}, we obtain that the two infinitesimal shift functions $\widehat\delta_x$ and $\widehat\delta_{x,y}$, seen as functionals of the density $\rho$, are related by
\begin{equation}\label{shift}
   \widehat\delta_{x,y}\big[\rho(z)\big]=\widehat\delta_x\left[\frac{\rho(z)}{\sum_l(x_l-y_l)}\right] + O(N^{-\infty}).
\end{equation}
This characterizes, at all orders in $N$, the infinitesimal difference of roots $x_j-y_j$.
We recall that, from \eqref{sum-x}, the sum
$\sum_l(x_l-y_l)$ is itself given by
\begin{align}
  \sum_l (x_l-y_l) &=\frac{\mathsf{k}_x-\mathsf{k}_y+\beta_x-\beta_y}{2(1-\eta)}+O(N^{-\infty})\label{diff2}\\
                            % =\frac{L(\beta_x-\beta_y)}{2(L-r)}
                              &=\frac{L(\mathsf{k}_x-\mathsf{k}_y)+2(\ell_x-\ell_y)}{2(L-r)}+O(N^{-\infty}),\label{dif-sum}
\end{align}
which also means that
\begin{equation}\label{id-om}
  e^{2\pi i(1-\eta)\sum_l (x_l-y_l)}
                             =e^{i\pi(\mathsf{k}_x -\mathsf{k}_y)}\, \frac{\omega_x}{\omega_y}+O(N^{-\infty}).
\end{equation}

In particular, the previous study enables us to identify (and count) the degenerate ground states associated with a given set of Bethe equations. We see from \eqref{shift} that two solutions $\{x\}$ and $\{y\}$ are different if and only if the total shift $\sum_l(x_l-y_l)$ is not an integer.
Hence, in this setting, the different degenerate ground states are completely determined by two quantum numbers $\mathsf{k}\in\mathbb{Z}/2\mathbb{Z}$ and $\ell\in  \mathbb{Z}/(L-r)\mathbb{Z}$.
If $L$ is odd, this gives $2(L-r)$ different values (modulo $1$) of \eqref{dif-sum}, corresponding to $2(L-r)$ different Bethe eigenstates. For even $L$, we only get $(L-r)$ different values of \eqref{dif-sum}, each of them being associated to two opposite values of $\omega$ according to the parity of $\mathsf{k}$ (see \eqref{id-om}).

%%%%%%%%%%%%%%%%%%%%%%%%%%%%%%
\section{The form factor in the thermodynamic limit}
\label{sec-ff-therm}

We now  study the thermodynamic limit of the renormalized form factor \eqref{magnetization} in the case where $\ket{\{u\},\omega_u}$ and $\ket{\{v\},\omega_v}$ are two ground states of the homogeneous model.

Let us set $x_j\equiv\tilde\eta u_j$, $y_j\equiv\tilde\eta v_j$, $j=1,\ldots,n$, and $\omega_x\equiv\omega_u$, $\omega_y\equiv\omega_v$.
We have seen in Section~\ref{sec-gs-th} that the Bethe roots $x_j$ (respectively $y_j$) for one of the ground states are completely determined by the data of two quantum numbers  $\mathsf{k} _x,\ell_x$ (respectively  $\mathsf{k}_y,\ell_y$). 
From now on, we simply denote $\ket{\mathsf{k}_x,\ell_x}\equiv\ket{\{u\},\omega_u}$ (respectively $\ket{\mathsf{k}_y,\ell_y}\equiv\ket{\{v\},\omega_v}$) the corresponding Bethe eigenstate, and $\ket{\psi_g^{(\mathsf{k}_x,\ell_x)}}$ (respectively $\ket{\psi_g^{(\mathsf{k}_y,\ell_y)}}$) the corresponding state renormalized to unity, i.e.
\begin{equation}
   \ket{\psi_g^{(\mathsf{k}_x,\ell_x)}}
   =\frac{\ket{\mathsf{k}_x,\ell_x}}{(\moy{\mathsf{k}_x,\ell_x\mid \mathsf{k}_x,\ell_x})^{1/2}},
   \qquad
   \ket{\psi_g^{(\mathsf{k}_y,\ell_y)}}
   =\frac{\ket{\mathsf{k}_y,\ell_y}}{(\moy{\mathsf{k}_y,\ell_y\mid \mathsf{k}_y,\ell_y})^{1/2}}.
\end{equation}

%So as to study the thermodynamic limit of the renormalized form factor \eqref{magnetization} in the case where $\ket{\{u\},\omega_u}$ and $\ket{\{v\},\omega_v}$ are two ground states of the homogeneous model, it will be convenient to rewrite the above determinant representation for $\mathbf{s}_m^z$  in terms of theta functions with quasi-period $\tilde\tau$ by means of Jacobi's imaginary transformation \eqref{jacobi-tr}.
%Let $x_j\equiv\tilde\eta u_j$, $y_j\equiv\tilde\eta v_j$, $j=1,\ldots,n$, and $\omega_x\equiv\omega_u$, $\omega_y\equiv\omega_v$. From the study of Section~\ref{sec-gs-th}, we know that these two ground states are completely characterized by two quantum numbers $\mathsf{k}_x,\ell_x$ (respectively  $\mathsf{k}_y,\ell_y$) and hence we use the notations $\ket{\mathsf{k}_x,\ell_x}$ and   $\ket{\mathsf{k}_y,\ell_y}$ (instead of $\ket{\{u\},\omega_u}$ and $\ket{\{v\},\omega_v}$).

In order to study how the determinant representation for $\mathbf{s}_m^z$ behaves in the thermodynamic limit, it is convenient to rewrite it in terms of theta functions with quasi-period $\tilde\tau$ by means of Jacobi's imaginary transformation \eqref{jacobi-tr}.
Let us first consider the quantity \eqref{ratio1} in the case $\gamma\not=0$.
We obtain
\begin{multline}\label{pol-theta}
%  \mathbf{s}_m^z(\{x\},\{y\})=
\frac{ \bra{\mathsf{k}_x,\ell_x}\, \sigma_m^z\, \ket{\mathsf{k}_y,\ell_y} } 
       { \moy{\mathsf{k}_y,\ell_y\mid \mathsf{k}_y,\ell_y}}
       =
 \bigg[ \frac{\omega_x}{\omega_y} e^{-2\pi i \eta \tilde\gamma}e^{-i\sum_{l=1}^n[p_0(x_l)-p_0(y_l)] } \bigg]^{m-1}
 \\
 \times
     \Bigg\{ \frac{1}{L} \sum_{s \in s_0 + \mathbb{Z}/L\mathbb{Z}} 
                \Big( \frac{\omega_y}{\omega_x} e^{2\pi i\eta \tilde\gamma } \Big)^{\! s}\frac{\theta_1(\tilde\eta s+\tilde\gamma) }{\theta_1(\tilde\eta s)} \Bigg\}
                     \left(-\tilde\eta\, \theta_1'(0)\, e^{-2\pi i \eta\tilde\gamma}\,\frac{\omega_x^2}{\omega_y^2}\right)^n
                                  \\
   \times 
     \prod_{k<l} \frac{\theta_1(y_k-y_l)}{\theta_1(x_k-x_l)}
     \frac{\det_n \big[ \widetilde{H}(\{x\},\{y\}) - 2 \widetilde{Q}(\{x\},\{y\}) \big] }
          { \det_n \big[ \widetilde\Phi (\{y\} ) \big] }     ,         
\end{multline}
with $\tilde\gamma=\sum_{j=1}^n (y_j-x_j)$ and
\begin{align}
  & \big[ \widetilde{H} \big]_{jk}
     = \frac{1}{\theta_1(\tilde\gamma)} \left\{ \frac{\theta_1(x_j-y_k+\tilde\gamma)}{\theta_1(x_j-y_k)}
                    -\frac{\omega_y}{\omega_x} e^{2\pi i\eta\tilde\gamma}
                    \frac{\theta_1(x_j-y_k+\tilde\gamma+\tilde\eta)}{\theta_1(x_j-y_k+\tilde\eta)}  \right\} 
        \prod_{l=1}^n \frac{\theta_1(x_l- y_k+\tilde\eta)}{\theta_1(y_l-y_k+\tilde\eta)}
       \nonumber\\
   &\hspace{1.2cm}
     -  \frac{1}{\theta_1(\tilde\gamma)}\left\{ \frac{\theta_1(x_j-y_k+\tilde\gamma)}{\theta_1(x_j-y_k)}
                    -\frac{\omega_x}{\omega_y} e^{-2\pi i\eta\tilde\gamma}
                     \frac{\theta_1(x_j-y_k+\tilde\gamma-\tilde\eta)}{\theta_1(x_j-y_k-\tilde\eta)}  \right\} 
                     \nonumber\\
    &\hspace{7.5cm}
    \times                 
        \Big(\frac{\omega_y}{\omega_x}\Big)^{\! 2} e^{4\pi i\eta\tilde\gamma}
        \prod_{l=1}^n \frac{\theta_1(x_l - y_k-\tilde\eta)}{\theta_1(y_l-y_k-\tilde\eta)}, 
     \label{mat-tH}\displaybreak[0]\\
 & \big[ \widetilde{Q} \big]_{jk}
     = \frac{1}{\theta_1(\tilde\gamma)}
        \left\{   \frac{\theta_1(x_j-\frac{\tilde\eta}{2}+\tilde\gamma)}{\theta_1(x_j-\frac{\tilde\eta}{2})}
                    -\frac{\omega_y}{\omega_x} e^{2\pi i\eta\tilde\gamma}\frac{\theta_1(x_j+\frac{\tilde\eta}{2}+\tilde\gamma)}{\theta_1(x_j+\frac{\tilde\eta}{2})}  \right\} 
        \prod_{l=1}^n \frac{\theta_1(x_l +\frac{\tilde\eta}{2})}{\theta_1(y_l+\frac{\tilde\eta}{2})},     
     \label{mat-tQ}\displaybreak[0]\\
 & \big[ \widetilde\Phi\big]_{jk}
      = -2\pi i \tilde\eta N \delta_{jk}\bigg\{\frac{ p_0'(y_j)}{2\pi}-\frac{1}{N} \sum_{l=1}^n K(y_j-y_l) \bigg\} - 2\pi i\tilde\eta K(y_j-y_k) +4\pi i \tilde\eta \eta. 
      \label{mat-tPhi}      
\end{align}

The determinant of the matrix $ \widetilde\Phi$ \eqref{mat-tPhi} is already in a convenient form for taking the thermodynamic limit. Using \eqref{sum-int}, \eqref{liebeq}, we get
\begin{equation}
   \big[ \widetilde\Phi \big]_{jk}
      =  -2\pi i \tilde\eta N \rho(y_k) \left\{ \delta_{jk} +\frac{1}{N} \frac{K(y_j-y_k)}{\rho(y_k)} -\frac{1}{N}\frac{2\eta}{\rho(y_k)} +O(N^{-\infty})\right\},
\end{equation}
so that the corresponding determinant can be written in terms of a Fredholm determinant 
%of kernel $K$ \eqref{K} 
in the thermodynamic limit:
\begin{equation}\label{Fred-norm}
  \det_n\big[ \widetilde\Phi(\{y\})\big]
   = (-2\pi i \tilde\eta N)^n \prod_{l=1}^n\rho(y_l)  \left\{ \det \big[ 1+ \widehat{K} -\widehat{V}_0\big]+O(N^{-\infty}) \right\}.
\end{equation}
Here $\widehat{K}$ and $\widehat{V}_0$ are integral operators acting on the interval $[-\frac12,\frac12]$, with respective kernels $K(y-z)$ given by \eqref{K}, and $V_0(y-z)=2\eta$.

The determinant appearing in the numerator of \eqref{pol-theta} can be transformed, similarly as what was done in \cite{IzeKMT99}, in a more suitable form for the thermodynamic limit. Using the results of Appendix~\ref{app-sp}, we have
\begin{multline}\label{id-detbis}
    (-1)^n\prod_{j<k}\frac{\theta_1(y_j-y_k)}{\theta_1(x_j-x_k)} \det_n \big[ \widetilde{H}(\{x\},\{y\}) - 2 \widetilde{Q}(\{x\},\{y\}) \big]  \\
    =\frac{1}{\theta_1(\tilde\gamma)}
    \Big\{ \det_n\!\big[ (\mathcal{H}-\mathcal{Q})(\{x\},\{y\})\big] 
              -\det_n\!\big[(\mathcal{H}+\mathcal{Q})(\{x\},\{y\})\big] \Big\},
\end{multline}
%
%
%\begin{multline}
%  \mathbf{s}_m^z(\{x\},\{y\}) 
%  = \bigg[ \frac{\omega_x}{\omega_y} e^{-2\pi i \eta \tilde\gamma}e^{-i\sum_{l=1}^n[p_0(x_l)-p_0(y_l)] } \bigg]^{m-1}
%     \Bigg\{ \frac{1}{L} \sum_{s \in s_0 + \mathbb{Z}/L\mathbb{Z}} 
%                \Big( \frac{\omega_y}{\omega_x} e^{2\pi i\eta \tilde\gamma } \Big)^{\! s}\frac{\theta_1(\tilde\eta s+\tilde\gamma) }{\theta_1(\tilde\gamma)\,\theta_1(\tilde\eta s)} \Bigg\}
%                                  \\
%   \times 
%     \big(\tilde\eta e^{-2\pi i \eta\tilde\gamma} \theta_1'(0)\big)^n
%         \frac{\det_n\! \big[ (\mathcal{M}- \mathcal{V})(\{x\},\{y\}) \big]- \det_n\! \big[ (\mathcal{M}+ \mathcal{V})(\{x\},\{y\}) \big]}
%           { \det_n\! \big[ \widetilde\Phi (\{y\} ) \big] }    ,   
%      \label{pol-theta-bis}
%\end{multline}
%
with
\begin{align}
  \big[ \mathcal{H}  \big]_{jk}
  & = \delta_{jk}\, 
     \frac{\prod_{l\not= j} \theta_1(y_j-y_l)}{\prod_{l=1}^n\theta_1(y_j-x_l)}
      \bigg\{ \prod_{l=1}^n  \frac{\theta_1(x_l-y_k+\tilde\eta)}{\theta_1(y_l-y_k+\tilde\eta)}
                 - \Big(\frac{\omega_y}{\omega_x}\Big)^{\! 2} e^{4\pi i\eta\tilde\gamma}\prod_{l=1}^n  \frac{\theta_1(x_l-y_k-\tilde\eta)}{\theta_1(y_l-y_k-\tilde\eta)} \bigg\}   \nonumber\\
  &\hspace{3.8cm}  + \frac{1}{\theta'_1(0)}\frac{\omega_y}{\omega_x} e^{2\pi i\eta\tilde\gamma} \left\{  \frac{\theta'_1(y_j-y_k+\tilde\eta)}{\theta_1(y_j-y_k+\tilde\eta)} 
                     -\frac{\theta'_1(y_j-y_k-\tilde\eta)}{\theta_1(y_j-y_k-\tilde\eta)} \right\}     , 
                    \nonumber \\
  &=   \delta_{jk}\, N\phi^{-1}_j(\{x\},\{y\}) \left\{ \phi_+(y_k|\{x\},\{y\})- \Big(\frac{\omega_y}{\omega_x}\Big)^{\! 2} e^{4\pi i\eta\tilde\gamma} \phi_-(y_k|\{x\},\{y\}) \right\} \nonumber\\
  &\hspace{3.8cm}  - \frac{2\pi i}{\theta'_1(0)}   \frac{\omega_y}{\omega_x} e^{2\pi i\eta\tilde\gamma}      K(y_j-y_k),        
       \label{mat-M}
\end{align}
and
\begin{equation}\label{mat-Q}
\big[ \mathcal{Q}  \big]_{jk}
  = \frac{\omega_y}{\omega_x} e^{2\pi i\eta\tilde\gamma} -\prod_{l=1}^n \frac{\theta_1(x_l+\frac{\tilde\eta}{2})\,\theta_1(y_l-\frac{\tilde\eta}{2})}
                                                  {\theta_1(y_l+\frac{\tilde\eta}{2})\, \theta_1(x_l-\frac{\tilde\eta}{2})}
  =\frac{\omega_y}{\omega_x} e^{2\pi i\eta\tilde\gamma} -e^{-i\sum_{l=1}^n[p_0(x_l)-p_0(y_l)] }.
\end{equation}
In \eqref{mat-M}, the factor $\phi_j(\{x\},\{y\})$ and the functions $\phi_\pm (y|\{x\},\{y\})$ are respectively defined as
\begin{align}
  &\phi_j(\{x\},\{y\}) =  N\frac{\prod_{l=1}^n\theta_1(y_j-x_l)}{\prod_{l\not= j} \theta_1(y_j-y_l)},
     \label{fct-phi}\\
  &\phi_\pm(y|\{x\},\{y\}) = \prod_{l=1}^n \frac{\theta_1(x_l-y\pm\tilde\eta)}{\theta_1(y_l-y\pm\tilde\eta)}.
     \label{fct-phi_eps}
\end{align}

The behavior of the matrix elements of $\mathcal{Q}$ can straightforwardly be evaluated in the thermodynamic limit. From \eqref{sum-px} and \eqref{id-om}, we obtain
%Using \eqref{Bethe-log} and \eqref{dif-sum}, it is easy to see that the difference of momentum between the two (quasi-)ground states is
%
%\begin{equation}
%   \sum_{l=1}^n[p_0(x_l)-p_0(y_l)] = 2\pi \sum_{l=1}^n(x_l-y_l)+O(N^{-\infty}).
%\end{equation}
%
%Hence, using \eqref{id-om}, we obtain that the elements of the matrix $\mathcal{Q}$ in the thermodynamic limit are given by
%
\begin{equation}
\big[ \mathcal{Q}  \big]_{jk}= \big[ (-1)^{\mathsf{k}}-1\big] e^{2\pi i \tilde\gamma} + O(N^{-\infty}),
\end{equation}
where we have set $\mathsf{k}=\mathsf{k}_y-\mathsf{k}_x$. It follows in particular that the form factor \eqref{magnetization} vanishes in the thermodynamic limit when $\mathsf{k}=0$.

The behavior of the functions $\phi_\pm(y|\{x\},\{y\})$ \eqref{fct-phi_eps}, conveniently rewritten in the form
\begin{equation}\label{phi_eps-log}
  \phi_\pm(y|\{x\},\{y\})=\exp\left\{ \sum_{l=1}^n\Big( \log\theta_1(x_l-y\pm\tilde\eta)
                                                             -\log\theta_1(y_l-y\pm\tilde\eta) \Big)\right\} ,
\end{equation}
can be evaluated by means of Proposition~\ref{prop-sum-int}.
Indeed, the functions $g^{\pm}:z\mapsto \log\theta_1(z-y\pm\tilde\eta)$ being $\mathcal{C}^\infty$ on $\mathbb{R}$ with a 1-periodic derivative, one can apply \eqref{sum-int2} to each of the sums in \eqref{phi_eps-log}, with corresponding constants given by
\begin{equation}
  c_{g^\pm}=\int_{-1/2}^{1/2} \frac{\theta'_1(z-y\pm\tilde\eta)}{\theta_1(z-y\pm\tilde\eta)} dz = \mp i\pi.
\end{equation}
One obtains
\begin{equation}\label{phi-pm}
   \phi_\pm(y|\{x\},\{y\}) = e^{\pm i\pi\tilde\gamma}+O(N^{-\infty}).
\end{equation}
So as to evaluate the factor $\phi$ \eqref{fct-phi}, let us define the function
\begin{equation}\label{fct-phi-Cinfty}
  \phi(y|\{x\},\{y\})=\prod_{l=1}^n \frac{\theta_1(x_l-y)}{\theta_1(y_l-y)}
     \left\{ \omega_y^2\, e^{4i\pi\eta \sum_{l} y_l  }\,
  \frac{\theta_1^N(y+\tilde\eta/2)}{\theta_1^N(y-\tilde\eta/2)}
  \prod_{l=1}^n \frac{\theta_1(y_l-y+\tilde\eta)}{\theta_1(y_l-y-\tilde\eta)},
  +1\right\}
\end{equation}
which is $\mathcal{C}^\infty$ and 1-periodic on $\mathbb{R}$. This function is such that
\begin{equation}\label{id-phij-fct}
   \phi(y_j|\{x\},\{y\})=-i\pi  \, \frac{\widehat\xi'_y(y_j)}{\theta_1'(0)}\, \phi_j(\{x\},\{y\}).
\end{equation}
Using the fact that the sum of the residues of the elliptic function
\begin{equation}
  h_\pm(z)= \frac{\theta_1(z-y-\tilde\gamma)}{\theta_1(z-y)}\prod_{l=1}^n \frac{\theta_1(z-x_l\pm\tilde\eta)}{\theta_1(z-y_l\pm\tilde\eta)}
\end{equation}
vanishes in an elementary cell, we obtain the following identity for the quantity $\phi_j$:
\begin{equation}
  \frac{1}{N}\sum_{j=1}^n\frac{\theta_1(y-y_j+\tilde\gamma\pm\tilde\eta)}{\theta_1(y-y_j\pm\tilde\eta)}\,\phi_j(\{x\},\{y\}) 
  =\theta_1(\tilde\gamma)\, \phi_\mp(y|\{x\},\{y\}).
\end{equation}
Rewriting the summand in terms of arguments of 1-periodic $\mathcal{C}^\infty$ functions evaluated at the ground state roots $y_j$ by means of \eqref{id-phij-fct}, and applying Proposition~\ref{prop-sum-int}, we obtain that, in the thermodynamic limit, the function $\phi$ \eqref{fct-phi-Cinfty} satisfies the integral equation
\begin{equation}
   -\frac{\theta_1'(0)}{2i\pi }\int_{-1/2}^{1/2} \frac{\theta_1(y-z+\tilde\gamma\pm\tilde\eta)}{\theta_1(y-z\pm\tilde\eta)}\,\phi(z)\, dz
   =e^{\mp i\pi\tilde\gamma} \,\theta_1(\tilde\gamma) +O(N^{-\infty}).
\end{equation}
The latter can easily be solved by Fourier transform and we get
\begin{equation}\label{lim-phi}
  -\frac{\theta_1'(0)}{2i\pi } \phi(y_j)=\phi_j\,\rho(y_j)+O(N^{-\infty})=\sin(\pi\tilde\gamma)\frac{\theta'_1(0)}{\pi}+O(N^{-\infty}).
\end{equation}
Gathering all these results, and using also \eqref{id-om}, we can therefore rewrite the elements of the matrix $\mathcal{H}(\{x\},\{y\})$ \eqref{mat-M} in the thermodynamic limit as
\begin{equation}
    \big[ \mathcal{H} \big]_{jk}
      =  -2\pi i N \frac{\rho(y_k)}{\theta'_1(0)} e^{2\pi i\tilde\gamma} \left\{ \delta_{jk} +\frac{(-1)^{\mathsf{k}}}{N} \frac{K(y_j-y_k)}{\rho(y_k)} +O(N^{-\infty})\right\}.
\end{equation}

Hence, the quantity \eqref{pol-theta} admits the following representation in terms of Fredholm determinants:
\begin{multline}
 %   \mathbf{s}_m^z(\{x\},\{y\}) = 
 \frac{ \bra{\mathsf{k}_x,\ell_x}\, \sigma_m^z \,\ket{\mathsf{k}_y,\ell_y} } 
       { \moy{\mathsf{k}_y,\ell_y\mid \mathsf{k}_y,\ell_y}}
% \frac{\bra{\{u\},\omega_u}\, \sigma_m^z \,\ket{\{v\},\omega_v} }{  \braket{ \{v\},\omega_v}{ \{v\},\omega_v } }
=
 (-1)^{\mathsf{k} (m-1)}
     \Bigg\{ \frac{1}{L} \sum_{s \in s_0 + \mathbb{Z}/L\mathbb{Z}} 
                \Big( \frac{\omega_y}{\omega_x} e^{2\pi i\eta \tilde\gamma } \Big)^{\! s}\frac{\theta_1(\tilde\eta s+\tilde\gamma) }{\theta_1(\tilde\eta s)\,\theta_1(\tilde\gamma)} \Bigg\}   
                             \\
     \times 
      \left( (-1)^{\mathsf{k}} \frac{\omega_x}{\omega_y} \right)^{\! n}\,
          \frac{\det \!\big[ 1+(-1)^{\mathsf{k}}\widehat{K}+\frac{1-(-1)^{\mathsf{k}}}{2}\widehat{V} \big]- \det \!\big[ 1+(-1)^{\mathsf{k}}\widehat{K}-\frac{1-(-1)^{\mathsf{k}}}{2}\widehat{V}   \big]}
            { \det \!\big[ 1+\widehat{K}-\widehat{V}_0 \big] }\\  + O(N^{-\infty}).      
\end{multline}
We recall that $\widehat{K}$ has kernel $K$ \eqref{K}, whereas $\widehat{V}_0$ and $\widehat{V}$ have respective constant kernels $2\eta$ and $\frac{i}{\pi}\theta'_1(0)$.

In its turn, the second ratio \eqref{ratio2} can be evaluated using the Bethe equations for $\{u\}$ and $\{v\}$, the estimations \eqref{phi-pm} and \eqref{lim-phi}, as well as the Fredholm determinant representation \eqref{Fred-norm}. It gives
\begin{equation}
\frac{ \moy{\mathsf{k}_y,\ell_y\mid \mathsf{k}_y,\ell_y}}
       { \moy{\mathsf{k}_x,\ell_x\mid \mathsf{k}_x,\ell_x}}
%\frac{ \braket{ \{v\},\omega_v}{ \{v\},\omega_v } }{\braket{ \{u\},\omega_u}{ \{u\},\omega_u } }
  =\left(\frac{\omega_y}{\omega_x}\right)^{\! 2n}+O(N^{-\infty}),
\end{equation} 
so that the conveniently renormalized form factor \eqref{magnetization} is simply given by
\begin{multline}\label{pol-theta-ter}
   \bra{\psi_g^{(\mathsf{k}_x,\ell_x)}}\,\sigma_m^z\,\ket{\psi_g^{(\mathsf{k}_y,\ell_y)}}
%    \mathbf{s}_m^z(\{x\},\{y\})
 =
    (-1)^{\mathsf{k} (m-1)}
     \Bigg\{ \frac{1}{L} \sum_{s \in s_0 + \mathbb{Z}/L\mathbb{Z}} 
                \Big( \frac{\omega_y}{\omega_x} e^{2\pi i\eta \tilde\gamma } \Big)^{\! s}\frac{\theta_1(\tilde\eta s+\tilde\gamma) }{\theta_1(\tilde\eta s)\,\theta_1(\tilde\gamma)} \Bigg\}   
                             \\
     \times 
          \frac{\det \!\big[ 1+(-1)^{\mathsf{k}}\widehat{K}+\frac{1-(-1)^{\mathsf{k}}}{2}\widehat{V} \big]- \det \!\big[ 1+(-1)^{\mathsf{k}}\widehat{K}-\frac{1-(-1)^{\mathsf{k}}}{2}\widehat{V}   \big]}
            { \det \!\big[ 1+\widehat{K}-\widehat{V}_0 \big] }  + O(N^{-\infty}).     
\end{multline}

The above Fredholm determinants can be computed by means of Fourier transform.
Indeed, as the kernel of the integral operator $\widehat{K}$ depends only on the difference of two variables, its eigenvalues correspond to the Fourier coefficients $k_m$ \eqref{km} of the function $K$ \eqref{K}. We obtain
\begin{align}
      \det \big[  1 + \widehat{K}-\widehat{V}_0 \big] 
       &= (2-2\eta) \prod_{m=1}^{+\infty}\left(1 + \tilde{q}^m\frac{1-\tilde{p}^m \tilde{q}^{-2m}}{1-\tilde{p}^m}\right)^{\! 2} 
       \nonumber\\
       &=2(1-\eta)\prod_{m=1}^{+\infty}\frac{(1+\tilde{q}^m)^2(1-\tilde{p}^m\tilde{q}^{-m})^2}{(1-\tilde{p}^m)^2},
\end{align}
whereas
\begin{align}       
     \det \bigg[ 1+(-1)^{\mathsf{k}}\widehat{K}\pm\frac{1-(-1)^{\mathsf{k}}}{2}\widehat{V}\bigg]
 %      &= \left( 1 +(-1)^{\delta n}\pm i\frac{1-(-1)^{\delta n}}{2\pi} \theta_1'(0)\right)
%       \nonumber\\
%       & \times
%        \prod_{m=1}^{+\infty}\left(1 +(-1)^{\delta n} \tilde{q}^m\frac{1-\tilde{p}^m\tilde{q}^{-2m}}{1-\tilde{p}^m}\right)^{\! 2} 
%        \nonumber\\
      &=  \left( 1 +(-1)^{\mathsf{k}}\pm i\frac{1-(-1)^{\mathsf{k}}}{2\pi} \theta_1'(0)\right)
       \nonumber\\
       & \times
        \prod_{m=1}^{+\infty}\frac{(1+(-1)^{\mathsf{k}}\tilde{q}^m)^2(1-(-1)^{\mathsf{k}}\tilde{p}^m\tilde{q}^{-m})^2}{(1-\tilde{p}^m)^2}.
\end{align}
Finally, the renormalized form factor \eqref{magnetization} is given as the following infinite product
\begin{multline}
 \bra{\psi_g^{(\mathsf{k}_x,\ell_x)}}\,\sigma_m^z\,\ket{\psi_g^{(\mathsf{k}_y,\ell_y)}}
 =
%    \mathbf{s}_m^z(\{x\},\{y\}) = 
 \frac{1-(-1)^{\mathsf{k}}}{2} (-1)^{m-1} 
 \prod_{m=1}^{+\infty}\frac{(1-\tilde{q}^m)^2(1+\tilde{p}^m\tilde{q}^{-m})^2}
  {(1+\tilde{q}^m)^2(1-\tilde{p}^m\tilde{q}^{-m})^2}
               \\
  \times
     \Bigg\{ \frac{i}{\pi(L-r)} \sum_{s \in s_0 + \mathbb{Z}/L\mathbb{Z}} 
                 e^{i\pi (2 \tilde\gamma-1)s } \, \frac{\theta_1(\tilde\eta s+\tilde\gamma)\, \theta'_1(0) }{\theta_1(\tilde\eta s)\,\theta_1(\tilde\gamma)} \Bigg\}    +O(N^{-\infty}).
                \label{pol-theta-result}
\end{multline}

The previous expression is a priori only valid in the case $\gamma\not=0$. One can perform a similar study in the two particular cases mentioned in Remark~\ref{rem-gamma=0}, i.e. for the mean value \eqref{mean-sgz} and the form factor \eqref{ff-sigmaz-bis} in the case $L$ even. One obtains that the former vanishes, whereas the latter can be written as
\begin{multline}
 \bra{\psi_g^{(0,\ell_x)}}\,\sigma_m^z\,\ket{\psi_g^{(1,\ell_x-L/2)}}
 =
%    \mathbf{s}_m^z(\{x\},\{y\}) = 
  (-1)^{m-1} 
 \prod_{m=1}^{+\infty}\frac{(1-\tilde{q}^m)^2(1+\tilde{p}^m\tilde{q}^{-m})^2}
  {(1+\tilde{q}^m)^2(1-\tilde{p}^m\tilde{q}^{-m})^2}
               \\
  \times
    \frac{i e^{-i\pi s_0} }{\pi(L-r)}\Bigg\{  \sum_{s=0}^{L-1} (-1)^s  \frac{\theta_1'\big(\tilde\eta (s_0+s)\big) }{\theta_1\big(\tilde\eta (s_0+s)\big)} -i\pi\eta L\Bigg\}    +O(N^{-\infty}).
                \label{pol-theta-result-bis}
\end{multline}

Note that we can rewrite a general representation valid for all cases by conveniently regularizing the representation \eqref{pol-theta-result} as
\begin{multline}
 \bra{\psi_g^{(\mathsf{k}_x,\ell_x)}}\,\sigma_m^z\,\ket{\psi_g^{(\mathsf{k}_y,\ell_y)}}
 =
%    \mathbf{s}_m^z(\{x\},\{y\}) = 
 \frac{1-(-1)^{\mathsf{k}}}{2} (-1)^{m-1} 
 \prod_{m=1}^{+\infty}\frac{(1-\tilde{q}^m)^2(1+\tilde{p}^m\tilde{q}^{-m})^2}
  {(1+\tilde{q}^m)^2(1-\tilde{p}^m\tilde{q}^{-m})^2}
               \\
  \times \lim_{\alpha\to 0}
     \Bigg\{ \frac{i}{\pi(L-r)} \sum_{s \in s_0 + \mathbb{Z}/L\mathbb{Z}} 
                 e^{i\pi (2 \tilde\gamma-1+2\eta\alpha)s } \, \frac{\theta_1(\tilde\eta s+\tilde\gamma+\alpha)\, \theta'_1(0) }{\theta_1(\tilde\eta s)\,\theta_1(\tilde\gamma+\alpha)} \Bigg\}    +O(N^{-\infty}).
                \label{pol-theta-alpha}
\end{multline}
We recall that, from \eqref{dif-sum}, the difference of roots $\tilde{\gamma}$ can be expressed in terms of the quantum numbers $\mathsf{k}=\mathsf{k}_y-\mathsf{k}_x$, $\ell=\ell_y-\ell_x$ as
\begin{equation}\label{tgamma}
   \tilde\gamma= \frac{L\mathsf{k}+2\ell}{2(L-r)}+O(N^{-\infty}),
\end{equation}
so that
\begin{multline}
  \bra{\psi_g^{(\mathsf{k}_x,\ell_x)}}\,\sigma_m^z\,\ket{\psi_g^{(\mathsf{k}_y,\ell_y)}}
  =  \frac{1-(-1)^{\mathsf{k}}}{2} (-1)^{m-1} 
     \prod_{m=1}^{+\infty}\frac{(1-\tilde{q}^m)^2(1+\tilde{p}^m\tilde{q}^{-m})^2}
  {(1+\tilde{q}^m)^2(1-\tilde{p}^m\tilde{q}^{-m})^2}
               \\
  \times
      \lim_{\alpha\to 0}
      \Bigg\{ \frac{i}{\pi(L-r)} \sum_{s \in s_0 + \mathbb{Z}/L\mathbb{Z}} 
      e^{2\pi i( \frac{r+2\ell}{2(L-r)}+\eta\alpha ) s}
     \frac{\theta_1\big(\tilde\eta s+\frac{L+2\ell}{2(L-r)}+\alpha\big)\, \theta'_1(0) }{\theta_1(\tilde\eta s)\,\theta_1\big(\frac{L+2\ell}{2(L-r)}+\alpha\big)} \Bigg\}    
  +O(N^{-\infty}).
  \label{pol-theta-result2}
\end{multline}
Note that this quantity depends only on the differences $\mathsf{k}$ and $\ell$ of the quantum numbers:
\begin{equation}
  \bra{\psi_g^{(\mathsf{k}_x,\ell_x)}}\,\sigma_m^z\,\ket{\psi_g^{(\mathsf{k}_y,\ell_y)}}
  \equiv \mathbf{s}_m^z(\mathsf{k},\ell) +O(N^{-\infty}).
\end{equation}
%

%%%%%%%%%%%%%%%%%%%%%%%%%%%%%%
\section{Spontaneous staggered polarizations of the CSOS model}
\label{sec-pol}

Let us consider the $2(L-r)$-dimensional subspace $\mathrm{Fun}(\mathcal{H}_g[0])$ of the space of states  $\mathrm{Fun}(\mathcal{H}[0])$ generated by all the ground states, i.e. by the Bethe eigenstates associated to the $2(L-r)$ largest (in magnitude) eigenvalues of the transfer matrix in the thermodynamic limit. A basis of this subspace is given by the normalized Bethe eigenstates   $\ket{\psi_g^{(\mathsf{k}_\alpha,\ell_\alpha)}}$, with $\mathsf{k}_\alpha\in\mathbb{Z}/2\mathbb{Z}$ and $\mathsf{\ell}_\alpha\in\mathbb{Z}/(L-r)\mathbb{Z}$.
This basis is not polarized, since the mean value of $\sigma_m^z$ in one of the Bethe ground states $\ket{\psi_g^{(\mathsf{k}_\alpha,\ell_\alpha)}}$ vanishes in the thermodynamic limit.
A polarized basis of $\mathrm{Fun}(\mathcal{H}_g[0])$ is instead given by
\begin{equation}\label{change-basis}
  \ket{\phi_g^{(\epsilon,\mathsf{t})}}
  =\frac{1}{\sqrt{2(L-r)}}\sum_{\mathsf{k}_\alpha=0}^1\sum_{\mathsf{\ell}_\alpha=0}^{L-r-1} 
  (-1)^{\mathsf{k}_\alpha\epsilon}e^{-i\pi\frac{r\mathsf{k}_\alpha+2\mathsf{\ell}_\alpha}{L-r}(\mathsf{t}+s_0)}\, \ket{\psi_g^{(\mathsf{k}_\alpha,\ell_\alpha)}},
\end{equation} 
with $\epsilon\in\{0,1\}$ and $\mathsf{t}\in \{0,1\ldots,L-r-1\}$.
Spontaneous staggered polarizations of the CSOS model are given as the mean values of the $\sigma_m^z$ operator in the polarized states \eqref{change-basis}.

\begin{rem}
It can be shown, for instance by considering the combinatorial formula for the Bethe states (see Theorem~5 of \cite{FelV96b}), that the combination \eqref{change-basis} tends, in the low-temperature limit $\tilde{p},\tilde{q}\to 0$, to one of the flat ground state configurations of the type $(a,a+1,a,a+1,\ldots)$ or $(a+1,a,a+1,a,\ldots)$ identified in \cite{PeaS89}.
More precisely, in the case $s_0=\frac{\tau}{2\eta}=-\frac{1}{2\tilde\eta}$, 
the state $\ket{\phi_g^{(\epsilon,\mathsf{t})}}$ with $\mathsf{t}=a-\lfloor \eta a\rfloor$ tends to the flat ground state configuration $(a,a+1,a,a+1,\ldots)$ if $\epsilon-\lfloor \eta a\rfloor$ is even, and to $(a+1,a,a+1,a,\ldots)$ if $\epsilon-\lfloor \eta a\rfloor$ is odd (here $\lfloor x \rfloor$ denotes the integer part of $x$).
\end{rem}

In fact, the matrix elements of the $\sigma_m^z$ operator in the states \eqref{change-basis} are given as
\begin{align}
   \bra{\phi_g^{(\epsilon_1,\mathsf{t}_1)}}\, \sigma_m^z\, \ket{\phi_g^{(\epsilon_2,\mathsf{t}_2)}}
   &=\frac{1}{2(L-r)}\sum_{\mathsf{k}_1,\mathsf{k}_2=0}^1\sum_{\mathsf{\ell}_1,\mathsf{\ell}_2=0}^{L-r-1} 
  (-1)^{\mathsf{k}_1\epsilon_1+\mathsf{k}_2\epsilon_2}
   \nonumber\\
  &\hspace{1cm}\times
    e^{i\pi\big[\frac{r\mathsf{k}_1+2\mathsf{\ell}_1}{L-r}(\mathsf{t}_1+s_0)-\frac{r\mathsf{k}_2+2\mathsf{\ell}_2}{L-r}(\mathsf{t}_2+s_0)\big]}\
    \bra{\psi_g^{(\mathsf{k}_1,\ell_1)}}\, \sigma_m^z\, \ket{\psi_g^{(\mathsf{k}_2,\ell_2)}}
  \nonumber\\
  &=  \frac{1}{2(L-r)}\sum_{\mathsf{k}_1,\mathsf{k}=0}^1\sum_{\mathsf{\ell}_1,\mathsf{\ell}=0}^{L-r-1} 
  (-1)^{\mathsf{k}_1(\epsilon_1-\epsilon_2)+\mathsf{k}\epsilon_2}
   \nonumber\\
     &\hspace{1cm}\times
     e^{i\pi\big[\frac{r\mathsf{k}_1+2\mathsf{\ell}_1}{L-r}(\mathsf{t}_1-\mathsf{t}_2)-\frac{r\mathsf{k}+2\mathsf{\ell}}{L-r}(\mathsf{t}_2+s_0)\big]} \
      \mathbf{s}_m^z(\mathsf{k},\ell)+O(N^{-\infty}),
\end{align}
in which we have set $\ell=\ell_2-\ell_1$ and $\mathsf{k}=\mathsf{k}_2-\mathsf{k}_1$.
We see from the sum over $\ell_1$ that this quantity is non-zero only if $\mathsf{t}_1=\mathsf{t}_2$, and then from the sum over $\mathsf{k}_1$ that one should also have $\epsilon_1=\epsilon_2$.
Hence, up to exponentially small corrections in the size of the system, the local operator $\sigma_m^z$ is diagonal in the basis \eqref{change-basis},
\begin{equation}
 \bra{\phi_g^{(\epsilon_1,\mathsf{t}_1)}}\, \sigma_m^z\, \ket{\phi_g^{(\epsilon_2,\mathsf{t}_2)}}
 = \delta_{\epsilon_1,\epsilon_2}\, \delta_{\mathsf{t}_1,\mathsf{t}_2} \,
 \bra{\phi_g^{(\epsilon_1,\mathsf{t}_1)}}\, \sigma_m^z\, \ket{\phi_g^{(\epsilon_1,\mathsf{t}_1)}}
 +O(N^{-\infty}),
\end{equation}
and the corresponding mean values are given as 
\begin{multline} 
 \bra{\phi_g^{(\epsilon,\mathsf{t})}}\, \sigma_m^z\, \ket{\phi_g^{(\epsilon,\mathsf{t})}}
   =\frac{i (-1)^{m-1+\epsilon} }{\pi (L-r)} \prod_{m=1}^{+\infty}\frac{(1-\tilde{q}^m)^2(1+\tilde{p}^m\tilde{q}^{-m})^2}
  {(1+\tilde{q}^m)^2(1-\tilde{p}^m\tilde{q}^{-m})^2}
  \\
  \times
     \lim_{\alpha\to 0}   \sum_{ \mathsf{\ell}=0}^{L-r-1} 
              \sum_{s=0}^{L-1} 
      e^{2\pi i( \frac{r+2\ell}{2(L-r)}+\eta\alpha ) (s-\mathsf{t})}\,
     \frac{\theta_1\!\big(\tilde\eta (s_0+s)+\frac{L+2\ell}{2(L-r)}+\alpha\big)\, \theta'_1(0) }
            {\theta_1\!\big(\tilde\eta(s_0+ s)\big)\,\theta_1\!\big(\frac{L+2\ell}{2(L-r)}+\alpha\big)}    
            +O(N^{-\infty}).\label{result1}
\end{multline}

The expression \eqref{result1} coincides with the one obtained by Date et al. in \cite{DatJKM90} in the case $r=1$, $L$ odd and $s_0=\frac{\tau}{2\eta}=-\frac{1}{2\tilde\eta}$.
Note however that the result \eqref{result1} can be presented in a much simpler form by means of the identities of Appendix~\ref{app-id-theta} for the sums of theta functions. Indeed, using respectively \eqref{id-sum1} to compute the sum over $\ell$, and then \eqref{id-sum2} to compute the sum over $s$, we obtain a simple combination of theta functions in which the $\alpha\to 0$ limit can be straightforwardly taken, so that
\begin{align}
\bra{\phi_g^{(\epsilon,\mathsf{t})}}\, \sigma_m^z\, \ket{\phi_g^{(\epsilon,\mathsf{t})}}
   &= (-1)^{m+\epsilon} \, \frac{i}{\pi}\,
     \frac{\theta'_1\big(0;\eta\tilde\tau\big)\,\theta_1\big(\tilde\eta \mathsf{t};(1-\eta)\tilde\tau\big)}
            {\theta_2\big(0;\eta\tilde\tau\big)\,\theta_2\big(\tilde\eta\mathsf{t};(1-\eta)\tilde\tau\big)}
     +O(N^{-\infty}) \label{result2}
            \\
   &=   (-1)^{m-1+\epsilon} \,
        \prod_{m=1}^{+\infty} \frac{(1-\tilde q^m)^2\, (1-\tilde p^m \tilde q^{-m-\mathsf{t}})}
                                                      {(1+\tilde q^m)^2\, (1+\tilde p^m \tilde q^{-m-\mathsf{t}})}  \,
        \prod_{m=0}^{+\infty}   \frac{ (1-\tilde p^m \tilde q^{-m+\mathsf{t}})}{  (1+\tilde p^m \tilde q^{-m+\mathsf{t}})}         \nonumber\\                                      
   &\hspace{8.5cm} +O(N^{-\infty}), \label{result3}
\end{align}
in which we have set $s_0=-\frac{1}{2\tilde\eta}.$

One recovers the low-temperature completely ordered limit by taking $|\tilde\tau|\to\infty$, i.e. $\tilde p,\tilde q\to 0$ in this expression.
If on the contrary one wants to study the critical limit $|\tau|\to \infty$, it is better to re-express the result \eqref{result2} by means of Jacobi's imaginary transformation as
\begin{equation}
\bra{\phi_g^{(\epsilon,\mathsf{t})}}\, \sigma_m^z\, \ket{\phi_g^{(\epsilon,\mathsf{t})}}
= (-1)^{m+\epsilon} \, \frac{i\tau}{\pi\eta}\,
    \frac{\theta'_1\big(0;\frac{\tau}{\eta}\big)\,\theta_1\big(\frac{ \eta\,\mathsf{t}}{1-\eta}; \frac{\tau}{1-\eta}\big)}
            {\theta_4\big(0;\frac{\tau}{\eta}\big)\,\theta_4\big(\frac{ \eta\,\mathsf{t}}{1-\eta}; \frac{\tau}{1-\eta}\big)}
     +O(N^{-\infty}).   \label{result4}
\end{equation}

%%%%%%%%%%%%%%%%%%%%%%%%%%%%%%%%%%%%%%%%%%%%%%%%%
%\section{Conclusion}
%\label{sec-concl}

%%%%%%%%%%%%%%%%%%%%%%%%%%%%%%%%%%%%%%%%%%%%%%%%%%%%%%%%%%%%%%%%%%%%%%%%%%%%%%%%%%%%%%%%%%%%%%%%%%%%%%
\section*{Acknowledgements}

V. T. is supported by CNRS.
We also acknowledge  the support from the ANR grant DIADEMS 10 BLAN 012004.
V. T. would  like to thank LPTHE (Paris VI University) for hospitality.

%%%%%%%%%%%%%%%%%%%%%%%%%%%%%%%%%%%%%%%%%%%%%%%%%%%%%%%%%%%%%%%%%%%%%%%%%%%%%%%%%%%%%%%%%%%%%%%%%%%%%%
\appendix

\section{Theta functions and useful identities}
\label{app-id-theta}

In this paper, $\theta_1(z;\tau)$ denotes the usual theta function with quasi-periods $1$ and $\tau$,
\begin{equation}\label{theta1}
  \theta_1(z;\tau)=-i\sum_{k=-\infty}^{\infty} (-1)^k e^{i\pi\tau (k+\frac12)^2} e^{2i\pi (k+\frac12)z},
  \qquad \Im\tau>0,
\end{equation}
which satisfies
\begin{equation}\label{periods}
   \theta_1(z+1;\tau)=-\theta_1(z;\tau), \qquad
   \theta_1(z+\tau;\tau)= -e^{-i\pi\tau}\, e^{-2\pi i z}\, \theta_1(z;\tau).
\end{equation}
We also denote
\begin{equation}
  \theta_2(z;\tau)=\theta_1\Big(z+\frac12;\tau\Big),
  \qquad
  \theta_4(z;\tau)=-i \, e^{\frac{i\pi\tau}{4}}\, e^{i\pi  z}\,\theta_1\Big(z+\frac{\tau}{2};\tau\Big).
\end{equation}

We recall Jacobi's imaginary transformation for the theta functions:
\begin{align}\label{jacobi}
   &\theta_1(z;\tau)
   = -i\, (-i\tau)^{-\frac12}\, e^{-i\pi \frac{z^2}{\tau}}\ \theta_1\Big(-\frac{z}{\tau}\, ;\, -\frac1\tau\,\Big),
   \\
   &\theta_2(z;\tau)
   =  (-i\tau)^{-\frac12}\, e^{-i\pi \frac{z^2}{\tau}}\ \theta_4\Big(-\frac{z}{\tau}\, ;\, -\frac1\tau\,\Big).
\end{align}

We also recall two useful summation identities (see for instance \cite{Liu12}):
\begin{align}
&\frac{1}{n}\sum_{\ell=0}^{n-1} e^{-2\pi i k \frac{\ell}{n} }\, 
  \frac{\theta_1\big(x+y+\frac{\ell}{n};\tau\big)\, \theta'_1\big(0;\tau\big)}
         {\theta_1\big(x ;\tau\big)\, \theta_1\big(y+\frac{\ell}{n};\tau\big)}
   =
    e^{2\pi i k y}\,
    \frac{\theta_1\big(x+ny+k\tau;n\tau\big)\, \theta'_1\big(0;n\tau\big)}
         {\theta_1\big(x+k\tau ; n\tau\big)\, \theta_1\big(ny;n\tau\big)},\label{id-sum1}
\\
  &\sum_{\ell=0}^{n-1} e^{2\pi i \frac{\ell}{n} x}\, 
  \frac{\theta_1\big(x+y+\frac{\ell}{n}\tau;\tau\big)\, \theta'_1\big(0;\tau\big)}
         {\theta_1\big(x ;\tau\big)\, \theta_1\big(y+\frac{\ell}{n}\tau;\tau\big)}
   =
    \frac{\theta_1\big(\frac{x}{n}+y;\frac{\tau}{n}\big)\, \theta'_1\big(0;\frac{\tau}{n}\big)}
         {\theta_1\big(\frac{x}{n} ;\frac{\tau}{n}\big)\, \theta_1\big(y;\frac{\tau}{n}\big)},
     \label{id-sum2}
  \end{align}
with $k\in\mathbb{Z}$.
These two identities are equivalent through Jacobi's imaginary transformation \eqref{jacobi} and quasi-periodicity property \eqref{theta1}.

%%%%%%%%%%%%%%%%%%%%%%%%%
\section{A determinant identity}\label{app-sp}

For two different sets of $n$ pairwise distinct variables $\{ x\}$ and $\{y\}$, and $\tilde\gamma=\sum_{j=1}^n(y_j-x_j)$, we consider the determinant
\begin{equation}
   \det_n \big[ \widetilde{H}_{\alpha}(\{x\},\{y\}) - 2 \widetilde{Q}_\beta(\{x\},\{y\}) \big] ,
\end{equation}
where $\widetilde{Q}_\beta$ is a  $n\times n$ rank-1 matrix and
\begin{multline}\label{mat-Halpha}
   \big[ \widetilde{H}_\alpha   \big]_{ij}
   =\frac{1}{\theta_1(\tilde\gamma)}\left\{
         \alpha_1 \frac{\theta_1(x_i-y_j+\tilde\gamma)}{\theta_1(x_i-y_j)}
        -\alpha_2 \frac{\theta_1(x_i-y_j+\tilde\gamma+\tilde\eta)}{\theta_1(x_i-y_j+\tilde\eta)}
        \right\}
     \prod_{l=1}^n\frac{\theta_1(x_l-y_j+\tilde\eta)}{\theta_1(y_l-y_j+\tilde\eta)} 
     \\
     -
     \frac{1}{\theta_1(\tilde\gamma)}\left\{
         \alpha_3 \frac{\theta_1(x_i-y_j+\tilde\gamma)}{\theta_1(x_i-y_j)}
        -\alpha_4 \frac{\theta_1(x_i-y_j+\tilde\gamma-\tilde\eta)}{\theta_1(x_i-y_j-\tilde\eta)}
        \right\} 
     \prod_{l=1}^n\frac{\theta_1(x_l-y_j-\tilde\eta)}{\theta_1(y_l-y_j-\tilde\eta)}  , 
\end{multline}
with $\alpha=(\alpha_1,\alpha_2,\alpha_3,\alpha_4)$ arbitrary.
For $t$ being an arbitrary complex parameter, we introduce the matrix $\mathcal{X}_t(\{x\},\{y\})$ given by
\begin{equation}\label{mat-X}
  \big[ \mathcal{X}_t   \big]_{jk} 
  = \frac{1}{\theta_1(t)} \, \frac{\prod_{l=1}^n \theta_1(x_k-y_l)}{\prod_{l\not= k} \theta_1(x_k-x_l)} \,
     \frac{\theta_1(y_j-x_k+t)}{\theta_1(y_j-x_k)} \, \frac{\theta_1(x_k)}{\theta_1(x_k-t)},
\end{equation}
with determinant
\begin{equation}\label{det-X}
  \det_n \big[\mathcal{X}_t (\{x\},\{y\}) \big] = (-1)^n\,\frac{\theta_1(\tilde\gamma+t)}{\theta_1(t)}
         \prod_{l=1}^n \frac{\theta_1(x_l)}{\theta_1(x_l-t)}
         \prod_{j<k} \frac{\theta_1(y_j-y_k)}{\theta_1(x_j-x_k)}.
\end{equation}
We have
\begin{equation}
  \det_n \big[ \widetilde{H}_{\alpha}(\{x\},\{y\}) - 2 \widetilde{Q}_\beta(\{x\},\{y\}) \big]
  = \frac{\det_n \big[ (\mathcal{X}_t \widetilde{H}_{\alpha}
                                  - 2 \mathcal{X}_t \widetilde{Q}_\beta )(\{x\},\{y\}) \big] }
             {  \det_n \big[\mathcal{X}_t (\{x\},\{y\}) \big] }.
\end{equation}

We can compute the product of matrices $\mathcal{X}_t \widetilde{H}_{\alpha}$ by means of the residue theorem applied to the functions
\begin{equation}\label{funct-g}
  g^{(j,k)}_{\eps}(z) = \frac{\theta_1(z-y_k+\tilde\gamma+\eps\tilde\eta)}{\theta_1(z-y_k+\eps\tilde\eta)}
    \prod_{l=1}^n\frac{\theta_1(z-y_l)}{\theta_1(z-x_l)} \,
    \frac{\theta_1(y_j-z+t)}{\theta_1(y_j-z)}\, \frac{\theta_1(z)}{\theta_1(z-t)},
\end{equation}
with $\eps\in\{0,+1,-1\}$ and $j,k=1,\ldots,n$.
These functions are doubly periodic of periods $1$ and $\tilde\tau$, they are therefore elliptic functions and the sum of their residues inside an elementary cell cancels. It leads to the identities
\begin{multline}\label{id-g}
   \sum_{b=1}^n \frac{\prod_{l=1}^n\theta_1(x_b-y_l)}{\prod_{l\not= b}\theta_1(x_b-x_l)}\,
   \frac{\theta_1(y_j-x_b+t)}{\theta_1(y_j-x_b)}\, \frac{\theta_1(x_b)}{\theta_1(x_b-t)}\cdot
   \frac{\theta_1(x_b-y_k+\tilde\gamma+\eps\tilde\eta)}{\theta_1(x_b-y_k+\eps\tilde\eta)}
   \\
   = -\theta_1(t)\, \frac{\theta_1(t-y_k+\tilde\gamma+\eps\tilde\eta)}{\theta_1(t-y_k+\eps\tilde\eta)}\,
        \prod_{l=1}^n\frac{\theta_1(t-y_l)}{\theta_1(t-x_l)} \, \frac{\theta_1(y_j)}{\theta_1(y_j-t)}
        \\
  -(1-\delta_{\eps,0})\,
    \theta_1(\tilde\gamma) \, \prod_{l=1}^n\frac{\theta_1(y_k-y_l-\eps\tilde\eta)}{\theta_1(y_k-x_l-\eps\tilde\eta)}\,
    \frac{\theta_1(y_j-y_k+t+\eps\tilde\eta)}{\theta_1(y_j-y_k+\eps\tilde\eta)}\, \frac{\theta_1(y_k-\eps\tilde\eta)}{\theta_1(y_k-t-\eps\tilde\eta)}
       \\
  +\delta_{j,k}\,\delta_{\eps,0}\, \theta_1(\tilde\gamma)\, \theta_1(t) \,
    \frac{\prod_{l\not= j} \theta_1(y_j-y_l)}{\prod_{l=1}^n\theta_1(y_j-x_l)}\,
    \frac{\theta_1(y_j)}{\theta_1(y_j-t)} .    
\end{multline}
Gathering all terms coming from the different versions of \eqref{id-g}, one obtains
\begin{equation}
  ( \mathcal{X}_t \widetilde{H}_{\alpha})  (\{x\},\{ y\}) 
  =  (\mathcal{H}_{\alpha,t} +2\mathcal{V}_{\alpha,t}) (\{x\},\{ y\}).
\end{equation}
Here
\begin{align}
   \big[ \mathcal{H}_{\alpha,t}  (\{x\},\{ y\}) \big]_{jk}
   &=
    \frac{\alpha_2}{\theta_1(t)} 
            \left[   \frac{\theta_1(y_j-y_k+t+\tilde\eta)}{\theta_1(y_j-y_k+\tilde\eta)} -1 \right]
                       \frac{\theta_1(y_k-\tilde\eta)}{\theta_1(y_k-t-\tilde\eta)}
                       \nonumber\\
     & -\frac{\alpha_4}{\theta_1(t)} \left[  \frac{\theta_1(y_j-y_k+t-\tilde\eta)}{\theta_1(y_j-y_k-\tilde\eta)} -1 \right] \frac{\theta_1(y_k+\tilde\eta)}{\theta_1(y_k-t+\tilde\eta)} 
                    \nonumber\\
    &\hspace{-3.2cm}
    +  \delta_{jk} \, \frac{\theta_1(y_j)}{\theta_1(y_j-t)}\, 
      \frac{\prod_{l\not= j} \theta_1(y_j-y_l)}{\prod_{l=1}^n\theta_1(y_j-x_l)}
      \bigg\{ \alpha_1\prod_{l=1}^n  \frac{\theta_1(x_l-y_k+\tilde\eta)}{\theta_1(y_l-y_k+\tilde\eta)}
                 -\alpha_3 \prod_{l=1}^n  \frac{\theta_1(x_l-y_k-\tilde\eta)}{\theta_1(y_l-y_k-\tilde\eta)} \bigg\}    ,    
      \label{mat-Mt}                            
\end{align}
and $\mathcal{V}_{\alpha,t}$ is a rank-1 matrix:
\begin{equation}
    \big[ \mathcal{V}_{\alpha,t}  (\{x\},\{ y\}) \big]_{jk}
   =
    \frac{1}{2\theta_1(t)}
     \bigg\{ \alpha_2\frac{\theta_1(y_k-\tilde\eta)}{\theta_1(y_k-t-\tilde\eta)}-\alpha_4\frac{\theta_1(y_k+\tilde\eta)}{\theta_1(y_k-t+\tilde\eta)}
     \bigg\}  
    +O(1),          
\end{equation}
where $O(1)$ stand for terms which remain finite in the limit $t\to 0$.

Recall that $t$ is an arbitrary parameter. We can therefore take the limit when it tends to $0$ to simplify the previous formula.
Using the fact that $\mathcal{V}_{\alpha,t}$ and $\mathcal{Q}_{\beta,t}\equiv\mathcal{X}_t \widetilde{Q}_\beta$ are of rank $1$, we have
\begin{equation}\label{limit1}
  \lim_{t\to 0} \, \theta_1(t)\, \det_n \big[ \mathcal{H}_{\alpha,t} + 2 (\mathcal{V}_{\alpha,t}-\mathcal{Q}_{\beta,t}) \big] 
  = \lim_{t\to 0}\, \theta_1(t)\,  \det_n  \mathcal{H}_{\alpha,t} 
      + 2 \lim_{t\to 0}\, \sum_{b=1}^n \det_n \mathcal{H}_{\alpha,t}^{(b)} , 
\end{equation}
with $\big[ \mathcal{H}_{\alpha,t}^{(b)} \big]_{jk} = \theta_1(t) \big[ \mathcal{V}_{\alpha,t} -\mathcal{Q}_{\beta,t}\big]_{jk}$ for $k = b$ and  $\big[ \mathcal{H}_{\alpha,t}^{(b)} \big]_{jk} = \big[ \mathcal{H}_{\alpha,t} \big]_{jk}$ otherwise.
The first limit in the right hand side of \eqref{limit1} is obviously zero, such that
\begin{align*} %\displaystyle
    \lim_{t\to 0} \, \theta_1(t)\, \det_n \big[ \mathcal{H}_{\alpha,t} + 2 (\mathcal{V}_{\alpha,t}-\mathcal{Q}_{\beta,t}) \big] 
    &= 2 \sum_{b=1}^n \det_n \big[ \mathcal{H}_\alpha^{(b)} \big] \nonumber\\
    &= \det_n\big[ \mathcal{H}_\alpha+(\mathcal{V}_\alpha-\mathcal{Q}_\beta)\big]
       -\det_n\big[\mathcal{H}_\alpha-(\mathcal{V}_\alpha-\mathcal{Q}_\beta)\big].
\end{align*}
Here we have defined
\begin{align}
  \big[\mathcal{H}_\alpha\big]_{jk} 
  &=\lim_{t\to 0} \big[ \mathcal{H}_{\alpha,t} \big]_{jk} \nonumber\\
  &= \delta_{jk} \, \frac{\prod_{l\not= j} \theta_1(y_j-y_l)}{\prod_{l=1}^n\theta_1(y_j-x_l)}
      \bigg\{ \alpha_1\prod_{l=1}^n  \frac{\theta_1(x_l-y_k+\tilde\eta)}{\theta_1(y_l-y_k+\tilde\eta)}
                 -\alpha_3 \prod_{l=1}^n  \frac{\theta_1(x_l-y_k-\tilde\eta)}{\theta_1(y_l-y_k-\tilde\eta)} \bigg\} 
                 \nonumber\\
   &\hspace{4mm}
   +\frac{1}{\theta_1'(0)} 
    \left[  \alpha_2 \frac{\theta_1'(y_j-y_k+\tilde\eta)}{\theta_1(y_j-y_k+\tilde\eta)} 
           -\alpha_4 \frac{\theta_1'(y_j-y_k-\tilde\eta)}{\theta_1(y_j-y_k-\tilde\eta)}  \right]  ,  \label{Halpha-new}              
\end{align}
\begin{align}
  &\big[\mathcal{V}_\alpha]_{jk} = \lim_{t\to 0} \theta_1(t) \big[\mathcal{V}_{\alpha,t} \big]_{jk}
    =\frac{\alpha_2-\alpha_4}{2}, \label{Valpha-new}\\
  & \big[ \mathcal{Q}_\beta \big]_{jk}= \lim_{t\to 0} \theta_1(t) \big[\mathcal{Q}_{\beta,t}\big]_{jk}, \label{mat-Qbeta}
\end{align}
and
\begin{equation} 
  \big[ \mathcal{H}_\alpha^{(b)} \big]_{jk} =   
  \begin{cases}
      \displaystyle \big[\mathcal{H}_\alpha]_{jk} & \text{if } k \neq b,\\
      \displaystyle \big[\mathcal{V}_\alpha-\mathcal{Q}_\beta\big]_{jk}  & \text{if } k=b.
  \end{cases}
\end{equation}
Hence,
\begin{multline}\label{id-det}
    \det_n \big[ \widetilde{H}_{\alpha}(\{x\},\{y\}) - 2 \widetilde{Q}_\beta(\{x\},\{y\}) \big]
   =\frac{(-1)^n}{\theta_1(\tilde\gamma)}\prod_{j<k}\frac{\theta_1(x_j-x_k)}{\theta_1(y_j-y_k)} \\
     \times
    \Big\{ \det_n\!\big[ (\mathcal{H}_\alpha+(\mathcal{V}_\alpha-\mathcal{Q}_\beta))(\{x\},\{y\})\big] 
              -\det_n\!\big[(\mathcal{H}_\alpha-(\mathcal{V}_\alpha-\mathcal{Q}_\beta))(\{x\},\{y\})\big] \Big\}.
\end{multline}

If we suppose moreover that the matrix $\widetilde{Q}_\beta$ is of the type
\begin{equation}
   \big[ \widetilde{Q}_\beta(\{x\},\{y\})   \big]_{ij}
   =\frac{1}{\theta_1(\tilde\gamma)}\left\{
         \beta_1 \frac{\theta_1(x_i-\frac{\tilde\eta}{2}+\tilde\gamma)}{\theta_1(x_i-\frac{\tilde\eta}{2})}
        -\beta_2 \frac{\theta_1(x_i+\frac{\tilde\eta}{2}+\tilde\gamma)}{\theta_1(x_i+\frac{\tilde\eta}{2})}
        \right\}
     \prod_{l=1}^n\frac{\theta_1(x_l+\frac{\tilde\eta}{2})}{\theta_1(y_l+\frac{\tilde\eta}{2})}   ,  
\end{equation}
with $\beta=(\beta_1,\beta_2)$ arbitrary, then we can compute the product $\mathcal{X}_t\widetilde{Q}_\beta$ similarly as above by means of the residue theorem applied to the elliptic function
\begin{equation}\label{funct-gQ}
  g^{(j)}_{\eps}(z) 
  = \frac{\theta_1(z-\eps\frac{\tilde\eta}{2}+\tilde\gamma)}{\theta_1(z-\eps\frac{\tilde\eta}{2})}
    \prod_{l=1}^n\frac{\theta_1(z-y_l)}{\theta_1(z-x_l)} \,
    \frac{\theta_1(y_j-z+t)}{\theta_1(y_j-z)}\, \frac{\theta_1(z)}{\theta_1(z-t)},
\end{equation}
with $\eps=\pm1$.
It leads to
\begin{equation}\label{Qbeta-new}
\big[ \mathcal{Q}_\beta \big]_{jk}
= \lim_{t\to 0} \theta_1(t) \big[\mathcal{X}_t\widetilde{Q}_\beta\big]_{jk}
= \beta_2-\beta_1\prod_{l=1}^n\frac{\theta_1(x_l+\frac{\tilde\eta}{2})\,\theta_1(y_l-\frac{\tilde\eta}{2})}{\theta_1(x_l-\frac{\tilde\eta}{2})\,\theta_1(y_l+\frac{\tilde\eta}{2})}.
\end{equation}

%%%%%%%%%%%%%%%%%%%%%%%%%%%%%%%%%%%%%%%%%%%%%%%%%%%%%%%%%%%%%%%%%%%%%%%%%%%%%%%%%%%%%%%%%%%%%%%%%%%%%%%%%%%%%%%%%%%%%%%%%%%%%%%%%%%%%%%%%%%%%%%%%%%%%%%%%

\providecommand{\bysame}{\leavevmode\hbox to3em{\hrulefill}\thinspace}
\providecommand{\MR}{\relax\ifhmode\unskip\space\fi MR }
\providecommand{\MRhref}[2]{%
  \href{http://www.ams.org/mathscinet-getitem?mr=#1}{#2}
}
\providecommand{\href}[2]{#2}

%\bibliographystyle{amsplain}
%\bibliographystyle{unsrt}
%\bibliographystyle{ieeetr}
%\bibliography{../biblio}

\end{document}